\documentclass[twoside,twocolumn,english,prb]{revtex4-1}

\usepackage[T1]{fontenc}
\usepackage[latin9]{inputenc}
\usepackage[a4paper]{geometry}
\usepackage{textcomp}
\usepackage{amsmath}
\usepackage{graphicx}
\usepackage{wasysym}
\usepackage{multirow}
\usepackage{xcolor}
\usepackage{ulem}

\makeatletter
\@ifundefined{date}{}{\date{}}
\usepackage{fancyhdr}
\makeatother

\usepackage{babel}
\begin{document}

\title{COMPETING PHASES AND CRITICAL BEHAVIOR IN THREE COUPLED SPINLESS
LUTTINGER LIQUIDS}

\author{S.Kundu$^{{1},{2}}$}
\author{and V.Tripathi$^{1}$}

\affiliation{$^{1}$Department of Theoretical Physics, Tata Institute of Fundamental Research, Homi Bhabha Road, Colaba, Mumbai 400005, India}
\affiliation{$^{2}$D\'epartement de physique and Institut quantique, Universit\'e de Sherbrooke,
Sherbrooke, Qu\'ebec, Canada J1K 2R1}
\date{\today}
\begin{abstract}
We study electronic phase competition in a system of three coupled
spinless Luttinger liquids using abelian bosonization, together with a perturbative renormalization group (RG) analysis. The scaling
procedure generates off-diagonal contributions to the
phase stiffness matrix, which require both rescaling as well as large
rotations of the fields. These rotations, generally non-abelian in nature, are important for correctly obtaining the dominant
electronic orders and critical behavior in different parameter regimes.
They generate a coupling between different interaction channels even at the
tree-level order in the coupling constant scaling equations. We study competing phases in this system, taking into account the aforementioned rotations, and determine
its critical behavior in a variety of interaction parameter regimes where perturbative RG is possible. The phase boundaries are found to be of the Berezinskii-Kosterlitz-Thouless (BKT) type, and 
we specify the parameter regimes where valley-symmetry breaking, chiral orders, and restoration of $C_{3}$ symmetry may be observed. We discuss experimental systems
where our approach and findings may be relevant. 
\end{abstract}
\maketitle

\section{\label{sec:introduction}introduction}

Coupled one-dimensional systems of interacting fermions appear in
diverse contexts. They have been used as building blocks for studying
higher-dimensional systems, such as cuprate high-temperature superconductors,\cite{article,PhysRevB.65.125106,PhysRevB.66.245109,PhysRevB.72.075126,PhysRevB.76.161101,PhysRevB.78.075124,PhysRevB.66.245106,PhysRevB.67.184517,PhysRevB.68.115104,PhysRevLett.83.2745}
due to the availability of controlled nonperturbative methods and
numerical techniques for analyzing them. They have also appeared in studies
of low-dimensional organic conductors,\cite{SUZUMURA200193} spin ladders,\cite{PhysRevLett.87.087205,PhysRevLett.86.1865,PhysRevB.79.205112,PhysRevB.76.054427,PhysRevB.53.8521,PhysRevB.50.252,Cabra2000,Allen_2001},
Mott insulating magnets, \cite{PhysRevB.96.205109} as well as artificially
manufactured structures (such as self-assembled transition metal nanowires
\cite{PhysRevB.85.115406}). Systems of three coupled Luttinger liquids
have, in general, received comparatively less attention than their two-coupled counterparts, but have been studied in the context of carbon nanotube systems,\cite{PhysRevB.74.085409,PhysRevB.82.155411} three-leg spin-tube models,\cite{Orignac1999, Sato2007,Charrier2010,Zhao2012,Fuji2014,Plat2015} coupled fermionic chains appearing in spin-ladder materials \cite{Cabra2000,Arrigoni1996,Arrigoni199691,Kimura1996,Kimura1998} and quasi-1D superconductors such as K$_{2}$Cr$_{3}$As$_{3}$\cite{PhysRevB.94.205129}.
The case of three spinless Luttinger liquids is especially interesting
since this is the simplest instance where orders such as chiral superconductivity
\cite{Kallin_2016} and chiral density wave can arise, which are not
possible in the case of Luttinger liquid systems with two or fewer
fermionic species. Experimentally, understanding the physics of three
coupled spinless Luttinger liquids may be useful in the context of multipocket
systems such as bismuth \cite{Behnia1729,Fauqu__2009,Kuchler2014,Li547,PhysRevB.79.081102,PhysRevB.79.241101,PhysRevB.79.245124,PhysRevB.80.075313,PhysRevB.84.115137,PhysRevLett.103.136803,Yang2010,Zhu14813}, graphite intercalates \cite{dresselhaus1981intercalation,Vogel1979} and  even the newly discovered heavy fermion superconductor UTe$_{2}$ \cite{Fujimori2019,Knebel2019,Ran2019,Ran684,Aoki2019,
Miyake2019,Tokunaga2019,Sundar2019,Metz2019,Ishizuka2019,
Braithwaite2019,Miao2020,Niu2019,Hutanu2019,Jiao2019,
Rann2019,Bae2019,Yarzhemsky2020}
in a strong magnetic field. In the quantum limit, these behave effectively as one-dimensional systems. 

Bosonization,\cite{rao2000bosonization,physik1998} together with a scaling treatment, has been a common method for studying the low-energy
properties of such systems. \cite{PhysRevB.65.125106, PhysRevB.60.2299,PhysRevB.76.161101,PhysRevB.78.075124,PhysRevB.66.245106,PhysRevB.67.184517,PhysRevB.68.115104,PhysRevB.74.085409,SUZUMURA200193,Tsukamoto2000,Itoi1999,Azaria1999,
Lee2004,Azaria19991} For coupled Luttinger liquid systems with three or more fermionic species, the scaling procedure generically introduces
off-diagonal corrections to the stiffness matrix $\widehat{K}$ in the quadratic part of the bosonized Hamiltonian (a sine-Gordon model): these corrections have largely been neglected in the existing analyses, \cite{PhysRevB.68.115104, PhysRevB.74.085409, PhysRevB.94.205129} and 
need to be taken into account.  They
carry information about the competition between different interaction channels, which in turn governs the electronic
phase competition and critical behavior in these systems. Although they have been introduced in a study involving two spinful coupled Luttinger liquids, \cite{PhysRevB.76.161101}  in the context of competing 
orders in cuprates, the specific nature of the interactions considered
there precludes the existence of chiral orders. On the other hand, the simplest situation
where such nontrivial corrections arise, is the case of three
coupled spinless Luttinger liquids. In this paper, we perform a one-loop RG analysis for such a system, which takes into account the effects of the off-diagonal
corrections by introducing large rotations
of the $\widehat{K}-$matrix and small renormalizations of the eigenvalues
of $\widehat{K}$. Of these two, the latter affects the scaling
dimensions of the interactions, while the former effectively rotates the bosonic fields, which affects the subsequent stages of the scaling.
From the solutions of the scaling equations, we identify the most
singular susceptibilities, corresponding to different order parameters,
which in turn determines the phase diagram. Also, from a numerical
scaling analysis of the RG equations, we obtain the critical behavior near the
phase transition points. 

Our main findings are as follows. We find that the fixed point behavior
is dependent both on the relative initial values of the coupling constants
and the Luttinger liquid parameter. This is a  situation qualitatively different from that of systems with two or less than two fermionic species (where such an interplay between different interaction channels does not appear) and is a direct consequence of the
rotations of the stiffness matrices introduced in our approach. Depending
upon the relative initial values of the couplings and the Luttinger
parameters, we identify the different instabilities in the particle-particle
and particle-hole channels and the nature of their transitions across
phase boundaries. Further, we obtain the conditions under which valley
symmetry breaking and intervalley orders may appear in both these
channels. The possibility of chiral orders is also discussed in this
context. 

Our calculations are expected to be relevant for understanding phase transitions
and critical phenomena in systems with multiple small
Fermi pockets (like graphite intercalates \cite{dresselhaus1981intercalation,Vogel1979},  bismuth \cite{Zhu14813,Yang2010,PhysRevLett.103.136803,PhysRevB.84.115137,PhysRevB.80.075313,PhysRevB.79.245124,PhysRevB.79.241101,PhysRevB.79.081102,Li547,Kuchler2014,Fauqu__2009,Behnia1729} and UTe$_{2}$ \cite{Fujimori2019,Knebel2019,Ran2019,Ran684,Aoki2019,
Miyake2019,Tokunaga2019,Sundar2019,Metz2019,Ishizuka2019,
Braithwaite2019,Miao2020,Niu2019,Hutanu2019,Jiao2019,
Rann2019,Bae2019,Yarzhemsky2020})
subject to quantizing magnetic fields, and cylindrical nanotubes at
high fields. \cite{PhysRevB.74.085409} In general, such an analysis is applicable for studies of competing
phases in three coupled one-dimensional systems where the instability occurs at
energy scales much smaller than the chemical potential. However, in
situations where the instabilities appear at higher energy scales,
other approaches such as the parquet renormalization group approach
\cite{furukawa1998truncation,honerkamp2001breakdown,nandkishore2012chiral}
are more suitable.

The rest of the paper is organized as follows. Sec-II introduces the
fermionic Hamiltonian with most generic interactions, describes the
bosonization procedure and presents the bosonized Hamiltonian.
Sec-III describes the renormalization group procedure used in our
analysis, which takes into account both the renormalizations of the
eigenvalues of $\widehat{K}$ as well as the large rotations of the
$\widehat{K}-$matrices. In Sec-IV, we introduce test vertices corresponding to different order parameter fields and study their evolution under the renormalization group, to determine the possible instabilities in different channels. Finally, Sec-V
presents a discussion of our results, conclusions and future directions. 

\section{\label{sec:interacting-model-and}interacting model and bosonization}
\begin{figure}
\begin{centering}
(a)\hspace{2mm}\includegraphics[width=0.9\columnwidth]{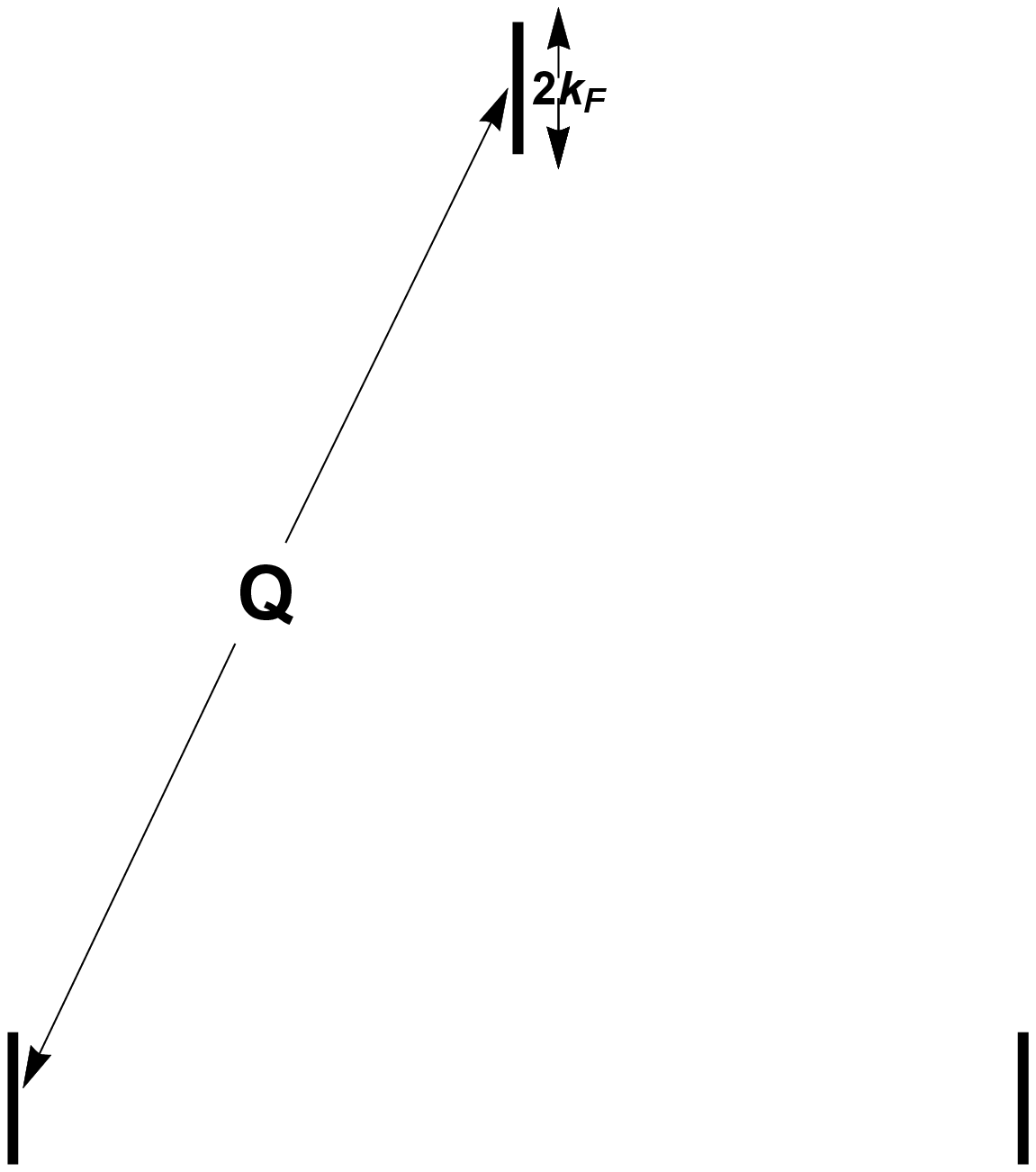}
\par\end{centering}
\vspace{0.5 in}
\begin{centering}
(b)\hspace{2mm}\includegraphics[width=0.9\columnwidth]{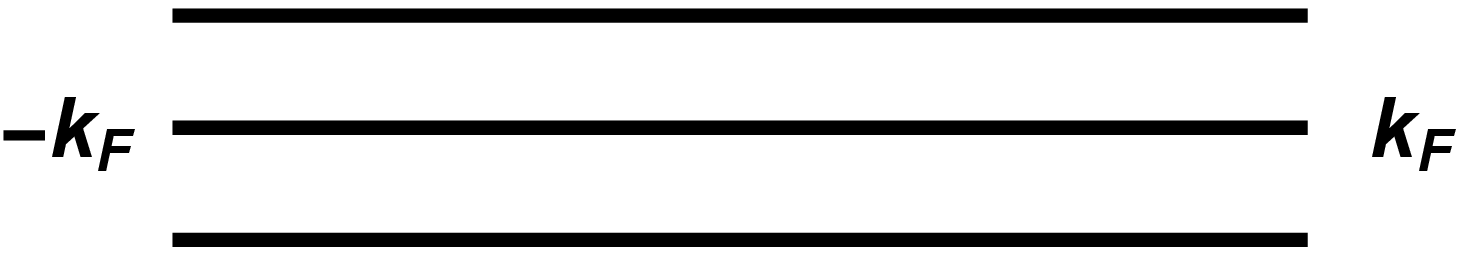}
\par\end{centering}
\caption{\label{fig:repr} The figure (a) above shows three small Fermi pockets, with Fermi momentum $k_{F}$, separated by a vector $Q$ in momentum space such that $Q>>k_{F}$, which is illustrative of the situation being considered in the present analysis. In contrast to this, in (b), each Fermi point comprises three flavors of fermions.}
\end{figure}

The fermionic Hamiltonian consists of two parts, 
\[
H=H_{0}+H_{int}
\]
where the noninteracting part is the three-band tightbinding model
describing electron hopping while the interacting part originates
from electron-electron interactions. The non-interacting
Hamiltonian in momentum space is given by 
\[
H_{0}=\sum_{km}\epsilon_{km}c_{km}^{\dagger}c_{km}
\]
where the band index $m=0,\pm1$, and $c_{km}(c_{km}^{\dagger})$
is the electron annihilation (creation) operator for the band $m$.
Near the Fermi points, the energy dispersion can be linearized as
$\epsilon_{km}=v_{Fm}(k-k_{Fm})$ where $v_{Fm}$ is the Fermi velocity
and $k_{Fm}$ is the Fermi momentum. We assume the Fermi momenta $k_{Fm}$ for the three
bands to be identical. 
We consider generic density-density type of interactions, and expand
the three spinless fermionic fields in the vicinity of the two Fermi
points. We are interested in situations that physically correspond
to partially filled bands, so that Umklapp scattering between the
two Fermi points for a given band is not relevant. However, since
we would like our model to be relevant for systems with multiple nested
Fermi pockets with a nesting vector equal to half a reciprocal lattice vector
(such as in the case of bismuth), we do allow the possibility of two-particle
Umklapp scattering between pockets, such that the total momentum transferred
corresponds to a reciprocal lattice vector. This situation is illustrated in figure 1(a) (In contrast, in figure 1(b), each Fermi point corresponds to three
flavors of fermions).

With these assumptions,
the interaction part of the Hamiltonian has the following form,
\begin{align}
H_{int} & =\sum_{p,m}(g_{1}^{(1)}\psi_{pm}^{\dagger}\psi_{\overline{p}\overline{m}}^{\dagger}\psi_{pm}\psi_{\overline{p}\overline{m}}+g_{1}^{(2)}\psi_{pm}^{\dagger}\psi_{\overline{p}\overline{m}}^{\dagger}\psi_{\overline{p}m}\psi_{p\overline{m}}\nonumber \\
 & +g_{1}^{(3)}\psi_{pm}^{\dagger}\psi_{p\overline{m}}^{\dagger}\psi_{\overline{p}m}\psi_{\overline{p}\overline{m}}+g_{1}^{(4)}\psi_{pm}^{\dagger}\psi_{p\overline{m}}^{\dagger}\psi_{pm}\psi_{p\overline{m}}\nonumber \\
 & +g_{2}^{(1)}\psi_{pm}^{\dagger}\psi_{\overline{p}\overline{m}}^{\dagger}\psi_{p\overline{m}}\psi_{\overline{p}m}+g_{2}^{(2)}\psi_{pm}^{\dagger}\psi_{\overline{p}\overline{m}}^{\dagger}\psi_{\overline{p}\overline{m}}\psi_{pm}\nonumber \\
 & +g_{2}^{(3)}\psi_{pm}^{\dagger}\psi_{p\overline{m}}^{\dagger}\psi_{\overline{p}\overline{m}}\psi_{\overline{p}m}+g_{2}^{(4)}\psi_{pm}^{\dagger}\psi_{p\overline{m}}^{\dagger}\psi_{p\overline{m}}\psi_{pm}\nonumber \\
 & +g_{3}^{(1)}\psi_{pm}^{\dagger}\psi_{\overline{p}m}^{\dagger}\psi_{p\overline{m}}\psi_{\overline{p}\overline{m}}+g_{3}^{(2)}\psi_{pm}^{\dagger}\psi_{\overline{p}m}^{\dagger}\psi_{\overline{p}\overline{m}}\psi_{p\overline{m}}\nonumber \\
 & +g_{3}^{(3)}\psi_{pm}^{\dagger}\psi_{pm}^{\dagger}\psi_{\overline{p}\overline{m}}\psi_{\overline{p}\overline{m}}+g_{3}^{(4)}\psi_{pm}^{\dagger}\psi_{pm}^{\dagger}\psi_{p\overline{m}}\psi_{p\overline{m}}\nonumber \\
 & +g_{4}^{(1)}\psi_{pm}^{\dagger}\psi_{\overline{p}m}^{\dagger}\psi_{pm}\psi_{\overline{p}m}+g_{4}^{(2)}\psi_{pm}^{\dagger}\psi_{\overline{p}m}^{\dagger}\psi_{\overline{p}m}\psi_{pm}\nonumber \\
 & +g_{4}^{(4)}\psi_{pm}^{\dagger}\psi_{pm}^{\dagger}\psi_{pm}\psi_{pm}),\label{eq:2-1}
\end{align}
where $p=1(-1)$ refers to right (left) moving fermions, and $m=0,1,-1$
denotes the bands and $\overline{m}\neq m$. The three bands are regarded
as identical, for simplicity. The above model is $C_{3}$ symmetric
under the  permutation of the three bands. To study the low-energy behavior, we shall
utilize the standard bosonization technique to analyze the continuum
fermion model. We now bosonize the fermionic model using the abelian
bosonization prescription,
\begin{equation}
\psi_{pm}=\frac{\eta_{pm}}{\sqrt{2\pi a}}\exp[ipk_{Fm}x]\exp[-ip\sqrt{\pi}\varphi_{pm}],\label{eq:2}
\end{equation}
where $k_{Fm}$ is the Fermi momentum for band $m$, $a$ is a cutoff
of the order of the lattice constant, and $p=1(-1)$ stands for the
$R(L)$ branch. The Majorana Klein factors $\eta_{R/Lm}$ satisfy 
\[
\{\eta_{Rm},\eta_{Rm^{\prime}}\}=2\delta_{mm^{\prime}}
\]
\[
\{\eta_{Lm},\eta_{Lm^{\prime}}\}=2\delta_{mm^{\prime}}
\]
\[
\{\eta_{Rm},\eta_{Lm^{\prime}}\}=0.
\]
We adopt the following convention for the Klein factors, following
Ref. \onlinecite{PhysRevB.94.205129}, 
\[
\eta_{mp}\eta_{\overline{m}p}=\eta_{0p}\eta_{mp}=imp,
\]
\[
\eta_{mp}\eta_{m\overline{p}}=\eta_{0p}\eta_{0\overline{p}}=ip,
\]
\[
\eta_{mp}\eta_{\overline{m}\overline{p}}=\eta_{0p}\eta_{m\overline{p}}=im,
\]
where $p,m=\pm1$. The chiral fields $\varphi_{pm}$ can be written
in terms of nonchiral fields $\phi_{m}$ and $\theta_{m}$ as $\varphi_{pm}=\phi_{m}-p\theta_{m}$,
and their gradients are proportional to the fermionic density and
current operators, respectively, i.e., 
\begin{align}
\nabla\phi_{m} & \propto\psi_{Rm}^{\dagger}\psi_{Rm}+\psi_{Lm}^{\dagger}\psi_{Lm}\nonumber \\
\nabla\theta_{m} & \propto\psi_{Rm}^{\dagger}\psi_{Rm}-\psi_{Lm}^{\dagger}\psi_{Lm}.\label{eq:3}
\end{align}
We collect all quadratic bosonic terms together, which we henceforth
call the ``noninteracting'' part. The rest consist of sine-Gordon
terms (see below) that we denote as interactions. 

We diagonalize the quadratic part of the bosonic Hamiltonian by transforming
to new bosonic fields $\widetilde{\phi_{i}}$ given by \cite{PhysRevB.94.205129}
\[
\left(\begin{array}{c}
\phi_{1}\\
\phi_{-1}\\
\phi_{0}
\end{array}\right)=\left(\begin{array}{ccc}
\frac{1}{\sqrt{2}} & \frac{1}{\sqrt{6}} & \frac{1}{\sqrt{3}}\\
-\frac{1}{\sqrt{2}} & \frac{1}{\sqrt{6}} & \frac{1}{\sqrt{3}}\\
0 & -\frac{2}{\sqrt{6}} & \frac{1}{\sqrt{3}}
\end{array}\right)\left(\begin{array}{c}
\widetilde{\phi}_{1}\\
\widetilde{\phi}_{-1}\\
\widetilde{\phi}_{0}
\end{array}\right),
\]
and likewise for the fields $\theta_{i}$. The ``noninteracting''
part of the Hamiltonian can then be written as 
\begin{equation}
H_{0}^{B}=\frac{1}{2}\int dx\sum_{\mu}v_{\mu}(K_{\mu}(\nabla\widetilde{\phi}_{\mu})^{2}+\frac{1}{K_{\mu}}(\nabla\widetilde{\theta}_{\mu})^{2}),\label{eq:3-1}
\end{equation}
where $\mu=0,1,-1$. Note our convention for $K_{\mu}$ differs from
the one commonly used in the literature, where $K_{\mu}^{-1}$ takes
the place of $K_{\mu}$. We have, for the bare couplings,
\begin{align*}
v_{\pm1}K_{\pm1} & =v_{F}-\frac{1}{2\pi}(G_{2}-G_{1})\equiv v_{\bot}K_{\bot}\\
v_{0}K_{0} & =v_{F}-\frac{1}{2\pi}(G_{2}^{0}-G_{1}^{0})\\
v_{\bot}= & \sqrt{(v_{F}-\frac{1}{2\pi}(G_{2}-G_{1}))(v_{F}+\frac{1}{2\pi}(G_{1}+G_{2}))}\\
K_{\bot}= & \sqrt{\frac{1-\frac{1}{2\pi v_{F}}(G_{2}-G_{1})}{1+\frac{1}{2\pi v_{F}}(G_{1}+G_{2})}}\\
v_{0}= & \sqrt{(v_{F}-\frac{1}{2\pi}(G_{2}^{0}-G_{1}^{0}))(v_{F}+\frac{1}{2\pi}(G_{1}^{0}+G_{2}^{0}))}\\
K_{0}= & \sqrt{\frac{1-\frac{1}{2\pi v_{F}}(G_{2}^{0}-G_{1}^{0})}{1+\frac{1}{2\pi v_{F}}(G_{1}^{0}+G_{2}^{0})}}
\end{align*}
where $G_{1}=g_{1}^{(4)}-g_{2}^{(4)}+g_{4}^{(4)}$, $G_{2}=-g_{1}^{(1)}+g_{2}^{(2)}+g_{4}^{(1)}-g_{4}^{(2)}$,
$G_{1}^{0}=-2g_{1}^{(4)}+2g_{2}^{(4)}+g_{4}^{(4)}$ and $G_{2}^{0}=2g_{1}^{(1)}-2g_{2}^{(2)}+g_{4}^{(1)}-g_{4}^{(2)}$.
The twofold degeneracy of the eigenvalues of the stiffness
matrix $\widehat{K}$ is a consequence of the $C_{3}$ rotational
symmetry of the quadratic part of the Hamiltonian. Following the strategy
of Ref. \onlinecite{PhysRevB.78.075124}, we study the scaling of
the quantities $K_{0,\bot}^{\phi}=v_{0,\bot}K_{0,\bot}$and
$K_{0,\bot}^{\theta}=\frac{v_{0.\bot}}{K_{0,\bot}}$, assuming
an initial condition $v_{0,\pm1}=1$. We now define new rescaled
fields $\widetilde{\psi}_{0,\pm1}=\sqrt{K_{0,\bot}^{\phi}}\widetilde{\phi}_{0,\pm1}$
and $\widetilde{\vartheta}_{0,\pm1}=\sqrt{K_{0,\bot}^{\theta}}\widetilde{\theta}_{0,\pm1}$.
Such a rescaling makes the stiffness matrix $\hat{K}$ proportional to the identity
matrix. During the RG process, we will find that small diagonal and off-diagonal corrections
are introduced to the stiffness matrix, and it has a real symmetric
form, that we denote by $Z_{\mu\nu}$. 

After bosonization, the ``interacting"
part of the bosonized Hamiltonian has the form of coupled sine-Gordon terms
\begin{equation}
H_{int}^{B}=\sum_{\alpha}g_{\alpha}\cos(a_{i}^{(\alpha)}\widetilde{\psi}_{i})+\sum_{\beta}g_{\beta}\cos(A_{i}^{(\beta)}\widetilde{\vartheta}_{i}),\label{eq:5-1}
\end{equation}
where $\alpha=1-3$, $7-9$ and $\beta=4-6$, and the coefficients,
\begin{align*}
a^{(1)} =\left(\begin{array}{ccc}
\frac{2\sqrt{2}\sqrt{\pi}}{\sqrt{K_{\bot}^{\phi}}}, & 0, & 0\end{array}\right),\\
a^{(2)} =\left(\begin{array}{ccc}
\frac{\sqrt{2}\sqrt{\pi}}{\sqrt{K_{\bot}^{\phi}}}, & \frac{\sqrt{6}\sqrt{\pi}}{\sqrt{K_{\bot}^{\phi}}}, & 0\end{array}\right),\\
a^{(3)} =\left(\begin{array}{ccc}
\frac{\sqrt{2}\sqrt{\pi}}{\sqrt{K_{\bot}^{\phi}}}, & -\frac{\sqrt{6}\sqrt{\pi}}{\sqrt{K_{\bot}^{\phi}}}, & 0\end{array}\right),\\
a^{(7)} =\left(\begin{array}{ccc}
0, & \frac{4\sqrt{\pi}}{\sqrt{6}\sqrt{K_{\bot}^{\phi}}} & \frac{4\sqrt{\pi}}{\sqrt{3}\sqrt{K_{0}^{\phi}}}\end{array}\right),\\
a^{(8)} =\left(\begin{array}{ccc}
\frac{\sqrt{2}\sqrt{\pi}}{\sqrt{K_{\bot}^{\phi}}}, & \frac{2\sqrt{\pi}}{\sqrt{6}\sqrt{K_{\bot}^{\phi}}}, & -\frac{4\sqrt{\pi}}{\sqrt{3}\sqrt{K_{0}^{\phi}}}\end{array}\right),\\
\end{align*}
\begin{align*}
a^{(9)} =\left(\begin{array}{ccc}
\frac{\sqrt{2}\sqrt{\pi}}{\sqrt{K_{\bot}^{\phi}}}, & -\frac{2\sqrt{\pi}}{\sqrt{6}\sqrt{K_{\bot}^{\phi}}}, & \frac{4\sqrt{\pi}}{\sqrt{3}\sqrt{K_{0}^{\phi}}}\end{array}\right),\\
A^{(4)} =\left(\begin{array}{ccc}
\frac{2\sqrt{2}\sqrt{\pi}}{\sqrt{K_{\bot}^{\theta}}}, & 0, & 0\end{array}\right),\\
A^{(5)} =\left(\begin{array}{ccc}
\frac{\sqrt{2}\sqrt{\pi}}{\sqrt{K_{\bot}^{\theta}}}, & \frac{\sqrt{6}\sqrt{\pi}}{\sqrt{K_{\bot}^{\theta}}}, & 0\end{array}\right),\\
A^{(6)} =\left(\begin{array}{ccc}
\frac{\sqrt{2}\sqrt{\pi}}{\sqrt{K_{\bot}^{\theta}}}, & -\frac{\sqrt{6}\sqrt{\pi}}{\sqrt{K_{\bot}^{\theta}}}, & 0\end{array}\right),
\end{align*}
where the effective couplings $g_{\alpha}(\alpha=1-9)$ are linear
combinations of the couplings $g_{i}^{(j)}$ (see Appendix \ref{app:A}). 
The validity of the perturbative RG analysis we shall perform below requires the coupling constants $g_\alpha$ to be small, and we assume this to be the case for the rest of the paper. However this limitation does not extend to the stiffnesses $K^{\phi,\theta},$ which may depart significantly from the noninteracting value $K^{\phi,\theta}=1,$ remaining within the purview of perturbative RG.
Indeed, given our motivation of understanding electronic phase competition in the quantum limit in low-carrier density (and consequently strongly correlated) semimetals such as bismuth, in the rest of the paper we will largely focus on regimes where the stiffnesses appreciably depart from unity.
Note that we allow the
possibility of the coupling constants in the sine-Gordon model to
break the $C_{3}$ permutation symmetry in the following analysis.
The same can also be done in the quadratic part and the two are equivalent. During the RG
procedure, the vectors $\widehat{a}$ and $\widehat{A}$, in general,
rotate and stretch.  The scaling dimensions for the interaction
terms in Eq. \ref{eq:5-1} depend on the values of the Luttinger parameters
$K_{0}^{\phi,\theta}$ and $K_{\perp}^{\phi,\theta}$, and in our
analysis, we only retain the most relevant interaction terms (with the smallest
scaling dimensions). This further reduces the number of parameters
we need to consider in our model.

\section{\label{sec:renormalization-group-analysis}renormalization-group
analysis}

The renormalization group follows the standard Wilsonian procedure
of elimination of fast degrees of freedom, restoration of the cutoff,
rescaling of the couplings and the renormalization of the fields.
This gives rise to off-diagonal
corrections in the stiffness matrices, which then take the form $Z_{\mu\nu}$. To keep the Gaussian
fixed point unchanged, we rotate the $Z_{\mu\nu}^{\theta,\phi}$
matrices, to diagonalize them, and then rescale the fields $\widetilde{\phi}_{i}$ or $\widetilde{\theta}_{i}$ (using the eigenvalues of these matrices) such that
the matrices become proportional to identity. Note that the above
rotation does not change the scaling dimensions of the sine-Gordon interaction
terms. Now, in the new basis obtained after the rotation and the subsequent rescaling
of the fields, we once again compute the one-loop corrections
and the resulting changes in the diagonal and off-diagonal elements of the stiffness matrices,
and repeat the aforementioned steps throughout the RG process. An
equivalent procedure has been followed in Ref. \onlinecite{PhysRevB.78.075124},
where, instead of keeping the Gaussian fixed point unchanged, the
fields are kept unchanged and the renormalization process leads to
rotations and stretching of eigenvalues of the $Z_{\mu\nu}^{\theta,\phi}$
matrices. We simplify our analysis by considering the anisotropic
limits $K_{\bot}^{\phi}\gg K_{0}^{\phi}$ or $K_{0}^{\phi}\gg K_{\bot}^{\phi}$,
which allows us to drop certain terms (which have higher scaling dimensions) in the interacting Hamiltonian
in Eq. \ref{eq:5-1} in each of these limits. However, the formulation may be readily extended to the most general case.
We note that the anisotropic limits $K_{\bot}^{\phi}\gg K_{0}^{\phi}$ or $K_{0}^{\phi}\gg K_{\bot}^{\phi}$ necessarily mean we are far from the noninteracting limit where $K^{\phi,\theta}\approx 1.$ Our remaining analysis thus corresponds to a strong coupling limit of the model. 
Below we discuss the results obtained by incorporating one-loop corrections
to the matrices $Z_{\mu\nu}^{\phi}$ and $Z_{\mu\nu}^{\theta}$ in
the two aforementioned anisotropic parameter regimes. At any given stage of the
RG, the matrix $Z_{\mu\nu}^{\phi}$, with the one-loop corrections
incorporated, is given by

\begin{widetext}

\begin{equation}
Z^{\phi}=\left(\begin{array}{ccc}
\frac{1}{2}+\sum_{\alpha}\frac{g_{\alpha}^{2}dy}{16\pi}((a_{1}^{(\alpha)})^{2}+(a_{-1}^{(\alpha)})^{2})(a_{1}^{(\alpha)})^{2} & \sum_{\alpha}\frac{g_{\alpha}^{2}dy}{16\pi}((a_{1}^{(\alpha)})^{2}+(a_{-1}^{(\alpha)})^{2})(a_{1}^{(\alpha)})(a_{-1}^{(\alpha)}) & 0\\
\sum_{\alpha}\frac{g_{\alpha}^{2}dy}{16\pi}((a_{1}^{(\alpha)})^{2}+(a_{-1}^{(\alpha)})^{2})(a_{1}^{(\alpha)})(a_{-1}^{(\alpha)}) & \frac{1}{2}+\sum_{\alpha}\frac{g_{\alpha}^{2}dy}{16\pi}((a_{1}^{(\alpha)})^{2}+(a_{-1}^{(\alpha)})^{2})(a_{-1}^{(\alpha)})^{2} & 0\\
0 & 0 & \frac{1}{2}
\end{array}\right).\label{eq:5-3}
\end{equation}

\end{widetext}

Note that the above matrix is block-diagonal - a consequence of the
nature of the interaction terms and/or approximations employed in
the parameter regimes considered in our analysis. While the corrections
accumulated are infinitesimal, the rotations involved in restoring
the matrices with off-diagonal contributions are finite rotations
which cannot be accounted for in the RG flow equations. In our approach,
we are always in the rotating frame, where these large rotations are
absent, and only small incremental changes to the components along
the field directions need to be tracked. These amount to slow changes
in the orientation and length, in the rotating frame, upon scaling.
In the limit where $K_{0}^{\phi}\ll K_{\perp}^{\phi}$, we find that
we only need to retain the couplings $g_{\alpha}(\alpha=1-3)$, based
on their lower scaling dimensions. In this case, we calculate one-loop
corrections to the $Z^{\phi}$ matrices due to the terms $g_{1}$,
$g_{2}$ and $g_{3}$ in the interaction Hamiltonian, and likewise,
to the $Z^{\theta}$ matrices due to the terms $g_{4}$, $g_{5}$
and $g_{6}$. The corresponding matrix turns out to be block-diagonal
due to the symmetry of the interaction terms in this regime. On the
other hand, in the limit where $K_{\perp}^{\phi}\ll K_{0}^{\phi}$,
only the couplings $g_{\alpha}(\alpha=7-9)$ need to be retained for
our analysis. Here we obtain one-loop corrections to the $Z^{\phi}$ matrices
arising from the couplings $g_{7}$, $g_{8}$ and $g_{9}$, and, once
again, to the $Z^{\theta}$ matrices due to the terms $g_{4}$, $g_{5}$
and $g_{6}$. In this case, the matrix $Z^{\phi}$ generally comprises nonzero corrections to every matrix element.
However, in the limit $K_{\perp}^{\phi}\ll K_{0}^{\phi}$,
we can drop certain terms and it reduces to a block-diagonal form
similar to Eq. \ref{eq:5-3} above with $\alpha=7-9$. 

In our analysis, we track the scaling equations for the interaction couplings, as well as the coefficients of the fields in the sine-Gordon terms. The eigenvalues of the matrix $Z_{\mu\nu}$ in Eq. \ref{eq:5-3}
above are denoted by $z_{1}$, $z_{-1}$ and $z_{0}$. We diagonalize the
matrix and then rescale the fields using these eigenvalues. At any given
stage of the RG flow, the coefficients of the fields in the cosine terms evolve in the manner $a_{i}^{(\alpha)}\rightarrow\frac{(Ra^{(\alpha)})_{i}}{\sqrt{z_{i}}}$,
where $R$ is the rotation which diagonalizes the matrix $Z_{\mu\nu}$.
We continue to denote the interaction terms as $g_{\alpha}\cos[\widehat{a}_{i}^{(\alpha)}\widetilde{\psi}_{i}]$
or $g_{\alpha}\cos[\widehat{A}_{i}^{(\alpha)}\widetilde{\vartheta}_{i}]$, and write down the RG equations for the coefficients ${a}_{i}^{(\alpha)}$, $A_{i}^{(\alpha)}$ and the couplings $g_{\alpha}$.
As an example, proceeding in incremental steps, the RG equations for the coefficients $a_{1}^{(1)}$
and $a_{-1}^{(1)}$ (corresponding to the coupling $g_{1}$) due to the rescaling process described above, are given by 
\begin{align}
\frac{da_{1}^{(1)}}{dy} & =-a_{1}^{(1)}\Lambda_{1}\nonumber \\
\frac{da_{-1}^{(1)}}{dy} & =-a_{-1}^{(1)}\Lambda_{-1}\label{eq:14-2}
\end{align}
 where $z_{1}=1/2+\Lambda_{1}dy$ and $z_{-1}=1/2+\Lambda_{-1}dy$, with $\Lambda_{1}$ and $\Lambda_{-1}$ depending upon all
 the coupling constants and the coefficients of all the fields in the sine-Gordon terms (see Appendix \ref{app:A}, for the explicit expressions of $\Lambda_{1}$ and $\Lambda_{-1}$). The leading corrections are quadratic in the coupling constants. This is not surprising since the RG equations of Eq. \ref{eq:14-2} essentially describe the renormalization of the stiffness constants $K^{\phi,\theta}$, which do not have $O(g)$ tree-level corrections.
The RG flow equations for the rest of the components $a_{i}^{(\alpha)}$
also behave in the same way. 

The tree-level contributions to the RG flows
for the sine-Gordon couplings $g_{\alpha}$ are obtained in terms of the scaling dimensions of the respective sine-Gordon terms, and the
one-loop contributions are obtained using the Operator Product Expansion (OPE).
The RG equations for the couplings $g_{\alpha},\alpha=1-3$ are
\begin{align}
\frac{dg_{1}}{dy} & =g_{1}(2-\frac{1}{4\pi}((a_{1}^{(1)})^{2}+(a_{-1}^{(1)})^{2}),\nonumber \\
 & +\frac{1}{8\pi}(a_{1}^{(2)}a_{1}^{(3)}+a_{-1}^{(2)}a_{-1}^{(3)})g_{2}g_{3},\nonumber \\
\frac{dg_{2}}{dy} & =g_{2}(2-\frac{1}{4\pi}((a_{1}^{(2)})^{2}+(a_{-1}^{(2)})^{2}),\nonumber \\
 & -\frac{1}{8\pi}(a_{1}^{(1)}a_{1}^{(3)}+a_{-1}^{(1)}a_{-1}^{(3)})g_{1}g_{3},\nonumber \\
\frac{dg_{3}}{dy} & =g_{3}(2-\frac{1}{4\pi}((a_{1}^{(3)})^{2}+(a_{-1}^{(3)})^{2}),\nonumber \\
 & -\frac{1}{8\pi}(a_{1}^{(1)}a_{1}^{(2)}+a_{-1}^{(1)}a_{-1}^{(2)})g_{1}g_{2}\label{eq:13-1}
\end{align}
The RG equations for the rest
of the couplings $g_{\alpha}(\alpha=4-9)$ are also easily obtained and have a similar form as Eq. \ref{eq:13-1}.

\subsection*{One-loop corrections to the RG equations}

The $O(g^2)$ one-loop (or OPE) contributions to the renormalization of the coupling constants $g_{\alpha}$ in Eq. \ref{eq:13-1} above are perturbatively smaller than the leading tree-level term. In contrast, the OPE contribution is the leading one in the RG equation for the coefficients of the fields in the cosine terms, given in Eq. \ref{eq:14-2}, which determine the scaling dimensions of the interaction terms. In general, one-loop corrections can have a significant effect on the RG flows when the tree-level term is small \textendash{} the usual motivation for considering higher order corrections in the perturbative RG. However, we found that if the initial values of the sine-Gordon couplings are small, and the initial stiffnesses are appreciably different from unity (reflecting the strongly correlated nature of our problem), the RG equations with or without the one-loop corrections generically give very similar solutions (see Fig. \ref{fig:tlol}).  If the initial scaling dimensions of the interaction terms are close to two (i.e. the tree-level contribution is small), or the bare values of the couplings are not sufficiently small (so that the one-loop and tree-level terms are comparable), then the one-loop terms need to be taken into account. This requires a separate, more detailed study and is not attempted in the present work.

We solve the coupled
differential equations 7 and 8 numerically and obtain the fixed-point values
for the various couplings $g_{\alpha}$ and the coefficients $a_{i}^{(\alpha)}.$
We consider weak repulsive interactions in every channel, and study
the nature of the RG flows as a function of the initial conditions
on the interactions and the value of the Luttinger liquid parameter
$K_{\bot}^{\phi,\theta}.$ In general, we find that the couplings
$g_{\alpha}$ either diverge or flow to zero in the course of the
RG flow. From Eq. \ref{eq:14-2} above, it is clear that the coefficients
$a_{1}^{(\alpha)}$ and $a_{-1}^{(\alpha)}$ obey different RG equations,
and show qualitatively different behavior as a function of the RG
flow parameter. In other words, the coefficients of the different
fields rescale differently in the course of the RG flow, following
the rotation of the stiffness matrix. 

\begin{figure}
\begin{centering}
\includegraphics[width=1.0\columnwidth]{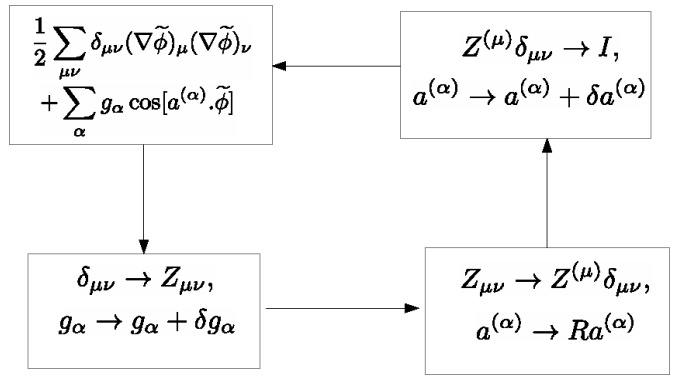}
\par\end{centering}
\caption{\label{fig:The-figure-shows}The figure shows a schematic illustration
of our renormalization group procedure. The stiffness matrix, which
is initially diagonal, develops off-diagonal corrections in the course
of the RG flow and takes the general form $Z_{\mu\nu}$. This matrix
is diagonalized, which leads to a rotation $R$ of the coefficients
$a^{(\alpha)}$ of the sine-Gordon interaction terms. The diagonal elements
are then absorbed in the respective sine-Gordon fields, which brings
the stiffness matrix back to unity, and leads to a rescaling of the
rotated coefficients $a^{(\alpha)}$. }
\end{figure}

\begin{figure}
\begin{centering}
\includegraphics[width=0.9\columnwidth]{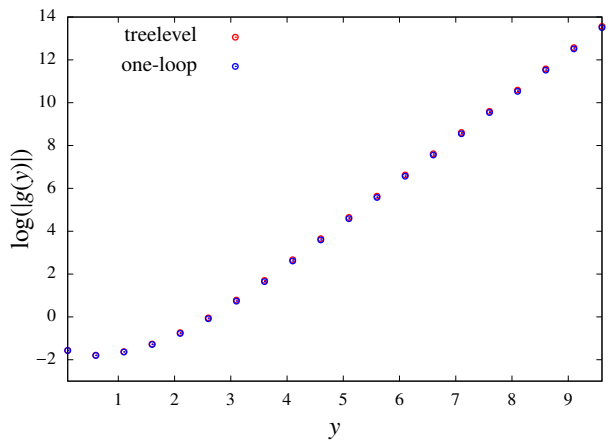}
\par\end{centering}
\caption{\label{fig:tlol}The figure compares 
the generic scaling behavior of the coupling $g_1$ with and without considering the effect of
the one-loop corrections in the scaling equations for the coupling constants. The parameters have been chosen such that the initial value of the tree-level term exceeds the one-loop contribution. The blue and red circles correspond to the cases with and without the one-loop contributions, respectively. The initial values of the couplings considered are $g_{1}=0.3$, $g_{2}=0.1$, $g_{3}=0.05$, and
the value of the Luttinger parameter $K_{\perp}^{\phi}=0.1.$ Clearly, the two sets of equations, with or without the one-loop contributions, give very similar results in this regime.}
\end{figure}

\begin{figure}
\begin{centering}
(a)\includegraphics[width=0.9\columnwidth]{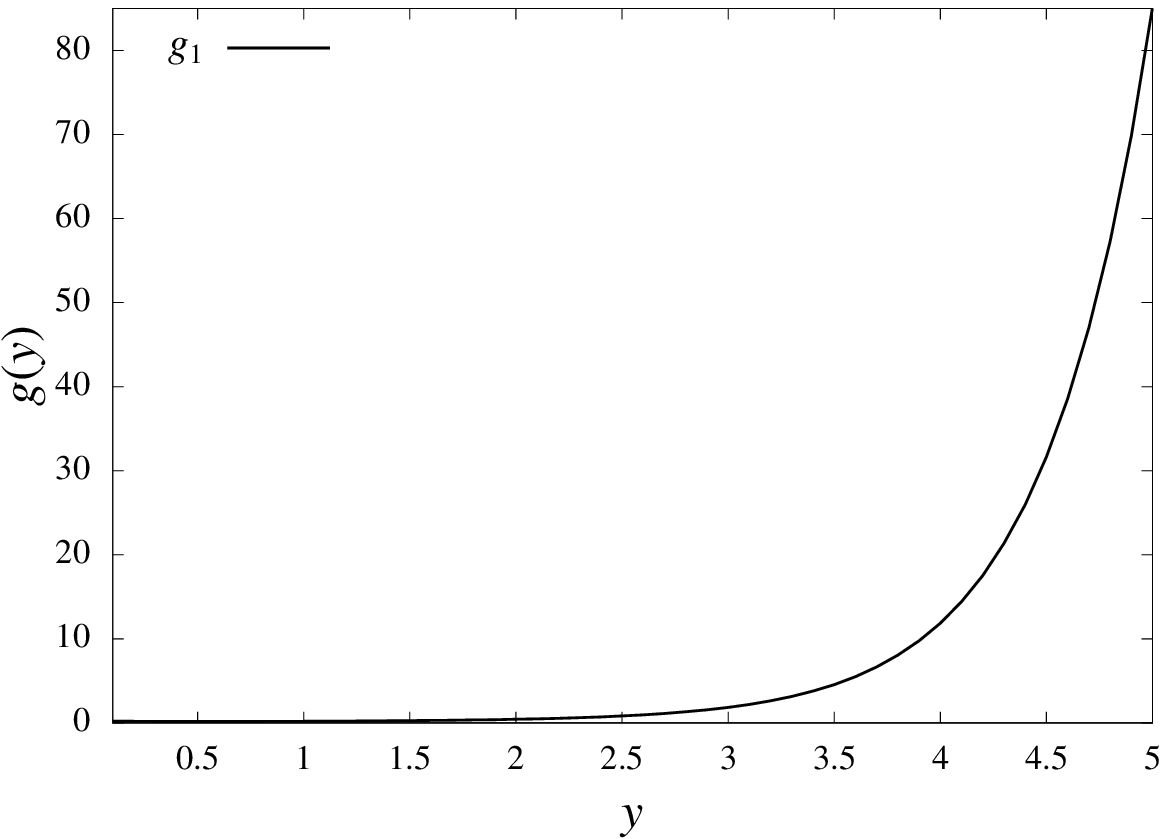}
\par\end{centering}
\begin{centering}
(b)\includegraphics[width=0.9\columnwidth]{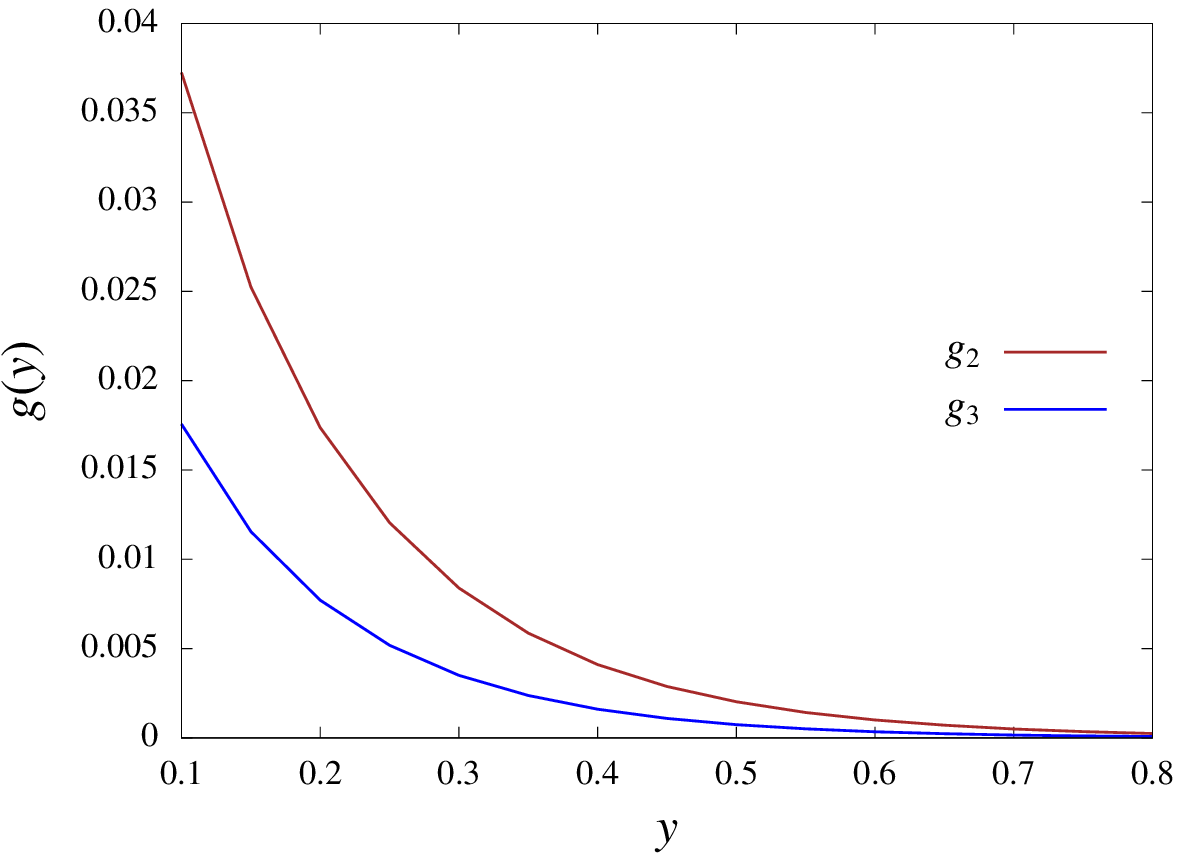}
\par\end{centering}
\caption{\label{fig:rgflows} The figure shows the RG flows for the couplings $g_{1}$,
$g_{2}$ and $g_{3}$ for the Luttinger parameter $K_{\perp}^{\phi}=0.1$ and initial conditions $g_{1}^{0}=0.3$, $g_{2}^{0}=0.1$, $g_{3}^{0}=0.05$. While $g_{1}$ grows monotonously (see (a))
under these conditions, $g_{2}$ and $g_{3}$ show a decline (see (b)). In general, any one or more of the couplings $g_{\alpha}$ may diverge, depending on the initial conditions chosen. The RG flows of the couplings $g_{i}$,$i=4-9$ behave in a manner qualitatively similar to that of $g_{1}$, $g_{2}$ and $g_{3}$.}
\end{figure}

\section{\label{sec:order-parameters-and}phase diagram and critical behavior}

\begin{figure}
\begin{centering}
\includegraphics[width=1.0\columnwidth]{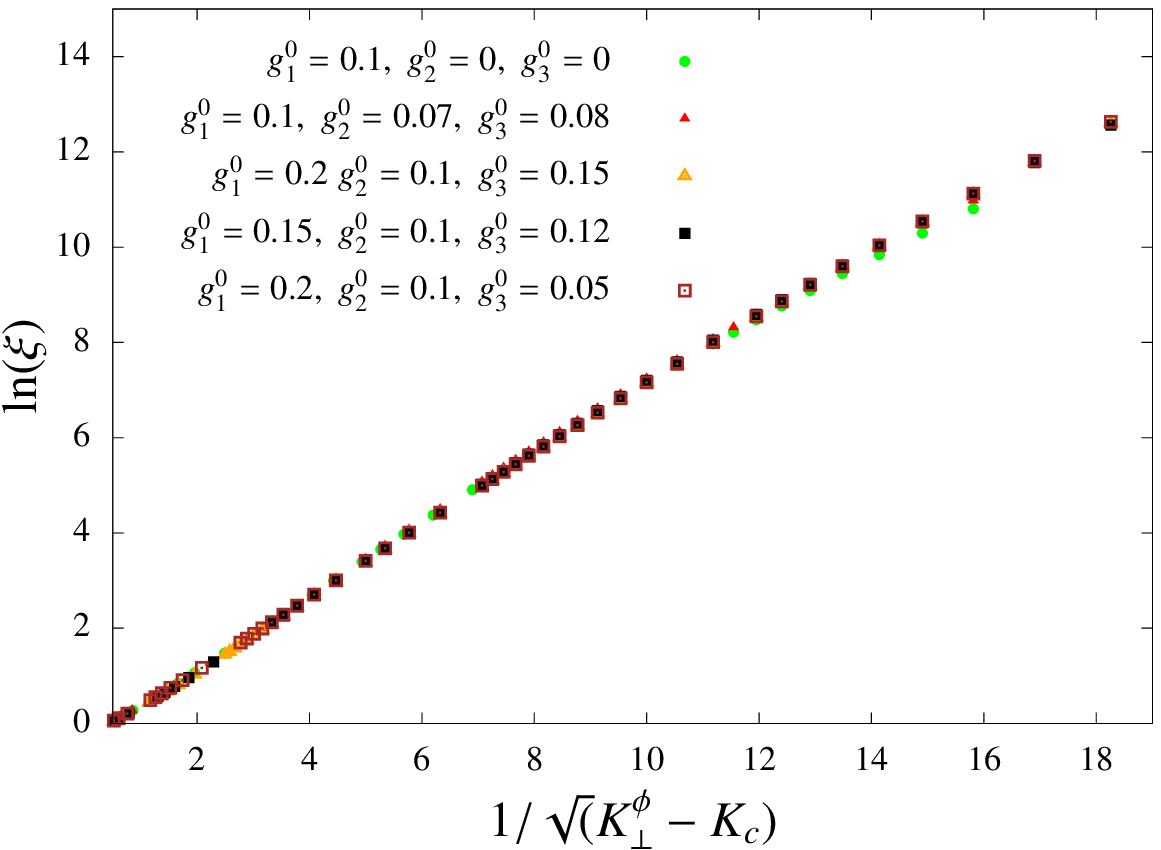}
\par\end{centering}
\caption{\label{fig:scalingcollapse}The figure shows a scaling collapse plot
of the RG flow parameter $y\sim\mathrm{ln}[\xi]$ (where $\xi$ is
the correlation length) as a function of $\frac{1}{\sqrt{K_{\bot}^{\phi}-K_{c}}}$.
$K_{c}$ denotes the critical value of the Luttinger liquid parameter
$K_{\bot}^{\phi}$, where the system undergoes a phase transition.
The plot shows results for five different sets of initial conditions
on the interactions, with one or more of the couplings $g_{\alpha}$
taking non-zero values initially, and indicates that the phase transitions
occuring in this system are continuous in nature and belong to the
BKT universality class. }
\end{figure}

The order parameters considered in our analysis are fermionic bilinear
operators characterized by chirality and band indices. There are two
classes of order parameters in our system. These are defined in the
particle-hole channel (density wave), \cite{PhysRevB.94.205129}
\begin{equation}
{\rm Re}[O_{ph}^{i0}]\propto\sum_{mm^{\prime}}\lambda_{mm^{\prime}}^{i}\psi_{Rm}^{\dagger}\psi_{Lm^{\prime}}+{\rm h.c,}\label{eq:6-1-1-1}
\end{equation}
and in the
particle-particle channel (superconductivity), 
\begin{equation}
{\rm Re}[O_{pp}^{i0}]\propto\sum_{mm^{\prime}}\lambda_{mm^{\prime}}^{i}\psi_{Rm}^{\dagger}\psi_{Lm^{\prime}}^{\dagger}+{\rm h.c},\label{eq:7}
\end{equation}
where $\lambda^{i} (i=1...8)$ correspond to the Gell-Mann matrices
(see Appendix \ref{app:B} for details), $\lambda^{0}$ denotes the 3x3 unit
matrix, and $\psi_{pm}$($\psi_{pm}^{\dagger}$) is the electron annihilation
(creation) operator with chirality $p$ and band $m$.
We follow the convention used by Ref.
\onlinecite{PhysRevB.94.205129} ; 
however, in both the Eqs. \ref{eq:6-1-1-1} and \ref{eq:7}, no spin
indices are present, due to the spinless nature of the fermions, indicated by the second index being $0$ for the order parameters.
Note that we consider ordered states arising from scattering or pairing
in opposite chiralities in this analysis, and we have checked that
equal-chirality interband pairing terms show a qualitatively similar
behavior. The order parameters in Eqs. \ref{eq:6-1-1-1} and \ref{eq:7}
above are expressed in terms of the bosonic fields. A total of
eighteen order parameters are obtained in the particle-hole and particle-particle
channels in the spinless case (see Appendix \ref{app:B} for expressions of the order parameters in terms of the bosonic fields). 

We now discuss the physical meaning of the electronic phases corresponding to the above order parameters. In the anisotropic strong coupling regime that we study (where the initial $K_{\perp}^{\phi}$ value is often far from unity and the initial $g_i$ are generically unequal), the phases that are obtained are typically associated with the breaking of valley permutation or bond permutation symmetries. However, we also find phases with the $C_{3}$ symmetry restored, not slaved to the initial conditions where this is explicitly broken (see Appendix B). Interband pairing 
in the particle-hole channel corresponds to a bond-ordered (BO) phase, while in the particle-particle channel it
gives rise to superconductivity at a finite wavevector (FFLO) equal to the separation between two small Fermi 
pockets in momentum space, $Q$. The intraband order parameters correspond to linear combinations of the fermionic bilinears on
the three different pockets. One of them is a symmetric linear combination ($s-$wave, denoted by SW) while the other two are nematic,
corresponding to angular momentum $l=2$ ($d-$wave order). If we ascribe the angular positions of the three patches in momentum space as $\delta=0$, $\delta=2\pi/3$ and $\delta=4\pi/3$,
the phases of the order parameters on the three valleys go either as $\cos(2\delta)$ or $\sin(2\delta)$, both of the $d-$wave type. 
It is also possible to have chiral orders, with phases going as $\exp(\pm i \delta)$, as a linear combination of nematic orders. These linear combinations are not
unique, and depending on the initial conditions, the actual order parameter may be some combination of these. 
Intraband pairing in the particle-hole channel has an ordering wavevector $2k_{F}$, much less than $Q$, and is generally incommensurate. Depending on the initial conditions,
the CDW (charge density wave) order could involve a linear combination of the CDW orders on the three different patches. If $C_{3}$ symmetry is not broken, then the orders may have $s-$wave (uniform CDW, denoted by UCDW), or a 
doubly degenerate $d-$wave symmetry ($d-$density wave). As was the case for superconductivity, the $d-$density wave order can be either nematic (denoted by NCDW) or chiral type (denoted by cCDW). The order parameters corresponding to different types of order are listed in Table \ref{tab:phases}. 

To study the dominant electronic orders, we introduce, in the disordered phase, test vertices corresponding to various order parameter fluctuations
and determine how they grow or shrink upon scaling. The evolution of any particular order parameter is governed by a certain combination of couplings, and the one 
with the smallest scaling dimension, such that the divergence is strongest upon scaling, is the dominant order. Those order parameters that initially have a large scaling dimension do not grow under scaling and correspond to short-range order. We also take into account the corrections to the scaling dimensions to leading order, $O(g)$ in the couplings, as these terms sometimes lead to shifts in the scaling dimensions of order parameters that have identical RG equations at the tree-level order, resulting in the lifting of degeneracies, with one of them becoming long-range ordered and the other short-range ordered (see Appendix \ref{app:B}). 

In order to determine the winning
order parameters, we consider the behavior of the couplings $g_{\alpha}$
and the corresponding coefficients of the fields $a_{i}^{(\alpha)}$ near
the fixed point of the RG for a given set of initial conditions and find
that both quantities play a crucial role in deciding the nature of
the dominant electronic orders. In some cases, we find that none of
the order parameters we studied grows under RG, implying the absence
of any quasi-long range ordered state despite the presence of interactions. Such situations also come up in the context of floating phases in coupled sine-Gordon models. The advantage of our method is that it not only gives us the dominant
order parameters, but also yields the scaling dimension at the fixed point which is essentially the exponent of power law correlations of the order parameter fields
in the quasi-long range ordered state. Later, we will show that the transitions, where they occur, belong to the Berezinskii-Kosterlitz-Thouless (BKT) universality
class, and that the correlation functions diverge upon 
approaching the critical point, in accordance with the BKT law. 

\begin{table*}
\begin{centering}
\begin{tabular}{|c|c|c|c|}
\hline 
\multicolumn{2}{|c|}{Type of order} & Order parameter & Name of order\tabularnewline
\hline 
\hline 
\multirow{4}{*}{Interband} & \multirow{2}{*}{p-p } & $O_{pp}^{10}$,$O_{pp}^{40}$,$O_{pp}^{60}$ & FFLO(wavevector $Q$)\tabularnewline
\cline{3-4} \cline{4-4} 
 &  & $O_{pp}^{20}$,$O_{pp}^{50}$,$O_{pp}^{70}$ & FFLO(wavevector $Q$)\tabularnewline
\cline{2-4} \cline{3-4} \cline{4-4} 
 & \multirow{2}{*}{p-h } & $O_{ph}^{10}$,$O_{ph}^{40}$,$O_{ph}^{60}$ & Bond order (BO)(wavevector $Q$)\tabularnewline
\cline{3-4} \cline{4-4} 
 &  & $O_{ph}^{20}$,$O_{ph}^{50}$,$O_{ph}^{70}$ & Bond order (BO)(wavevector $Q$)\tabularnewline
\hline 
\multirow{4}{*}{Intraband} & \multirow{2}{*}{p-p } & \multirow{2}{*}{$O_{pp}^{00}$,$O_{pp}^{30}$,$O_{pp}^{80}$} & \multirow{2}{*}{$s-$wave (SW), Nematic $d-$wave, Chiral $d-$wave}\tabularnewline
 &  &  & \tabularnewline
\cline{2-4} \cline{3-4} \cline{4-4} 
 & \multirow{2}{*}{p-h } & \multirow{2}{*}{$O_{ph}^{00}$,$O_{ph}^{30}$,$O_{ph}^{80}$} & \multirow{2}{*}{Uniform(U) CDW , nematic (N) $d-$CDW , chiral (c) $d-$CDW }\tabularnewline
 &  &  & \tabularnewline
\hline 
\end{tabular}
\par\end{centering}
\caption{\label{tab:phases} Table showing electronic phases corresponding to each of the order parameters considered in our analysis. Here particle-particle (p-p) refers to superconductivity, while particle-hole (p-h) refers to density wave orders. Interband pairing between different
pairs of bands in the particle-hole channel leads to bond order (denoted by BO) while
the corresponding pairing in the particle-particle channel leads to
a finite-momentum pairing (denoted by FFLO) state with the wavevector $Q$, equal to the separation between two small Fermi pockets in momentum space. Intraband pairing can correspond
to a situation with different phases on different Fermi pockets and
lead to uniform charge density wave (denoted by UCDW) or nematic $d-$density wave
order (denoted by NCDW) in the particle-hole channel, and $s-$wave or nematic $d-$wave
superconductivity in the particle-particle channel. In the case where
these different order parameters are degenerate, a combination of
them which is chiral in nature gives rise to the lowest
energy configuration. In such a situation, a
chiral $d-$density wave (denoted by cCDW) or chiral $d-$wave superconductivity can
be realized. Despite choosing initial conditions that generically break $C_3$ permutation symmetry, one nevertheless finds that in some parameter regimes (see text, Fig. \ref{fig:pd}), phases with the $C_{3}$ symmetry restored, such as the chiral orders, are dominant.}

\end{table*}

We classify the nature of the dominant orders in different parameter regimes depending upon the relative sizes of $K_{\bot}^{\phi}$ and  $K_{0}^{\phi}$, considering the two broad classes of parameters, $K_{0}^{\phi}\gg K_{\perp}^{\phi}$ and $K_{\perp}^{\phi}\gg K_{0}^{\phi}$. Clearly, this implies some $K_{\perp}^{\phi}$ values must necessarily take values far from the noninteracting point $K_{\perp}^{\phi}=1,$ i.e., we are in a strong-correlation regime that is nevertheless accessible by perturbative RG. Within each of these classes, we further examine situations with either $K_{0}^{\phi}\gg1$ or $K_{0}^{\phi}\ll1$.
The case with $K_{0}^{\phi}\sim1$, involving a competition between different types of orders, depending upon the initial conditions, requires a more detailed study,
and has not been addressed here. In the regime where $K_{0}^{\phi}\gg K_{\perp}^{\phi}$
and $K_{0}^{\phi}\ll1$, the dominant instabilities are found in the
intraband particle-particle channel. Similarly, in the regime where
$K_{\perp}^{\phi}\gg K_{0}^{\phi}$ and $K_{0}^{\phi}\gg1$, the dominant
instabilities are found in the intraband particle-hole channel. Note that in these two parameter regimes, $K_{\perp}^{\phi}$ is automatically constrained to be numerically very small or very large. We
now consider the remaining two cases, which allow us to tune $K_{\bot}^{\phi}$
over a wide range of values, giving rise to both intraband and interband orders. 

We find that for $K_{0}^{\phi}\gg K_{\perp}^{\phi}$ and $K_{0}^{\phi}\gg1$,
the particle-hole orders are more relevant than the particle-particle orders, due to smaller scaling
dimensions of the corresponding order parameters, and for $K_{\perp}^{\phi}\gg K_{0}^{\phi}$
and $K_{0}^{\phi}\ll1$, the particle-particle orders are likewise found to
be more important. Within the regimes considered by us,
the phase diagram is affected primarily by two factors: the magnitude
of the Luttinger liquid parameter $K_{\bot}^{\phi}$ and the set of
initial conditions considered for the interactions $g_{\alpha}$. The nature of the phase transitions is studied using a numerical scaling analysis. The scaling of the correlation length $\xi$ at the critical point
is determined by identifying the characteristic scale $y$ where the
couplings $g_{\alpha}(y)$ cross a designated value $\apprge1$. We
obtain continuous transitions as a function of $K_{\bot}^{\phi}$,
belonging to the Berezinskii-Kosterlitz-Thouless (BKT) universality
class, which is confirmed by demonstrating the universal BKT scaling
collapse for the behavior of the correlation length close to the critical
point (see Fig. \ref{fig:scalingcollapse}). Note that the critical
value $K_{c}$ of the Luttinger parameter $K_{\bot}^{\phi}$ is different
for different initial conditions on the couplings $g_{\alpha}$, as
shown in Fig. \ref{fig:scalingcollapse}, each of which give rise to the same
critical behavior. 


\begin{figure*}
\begin{centering}
\includegraphics[width=1.6\columnwidth]{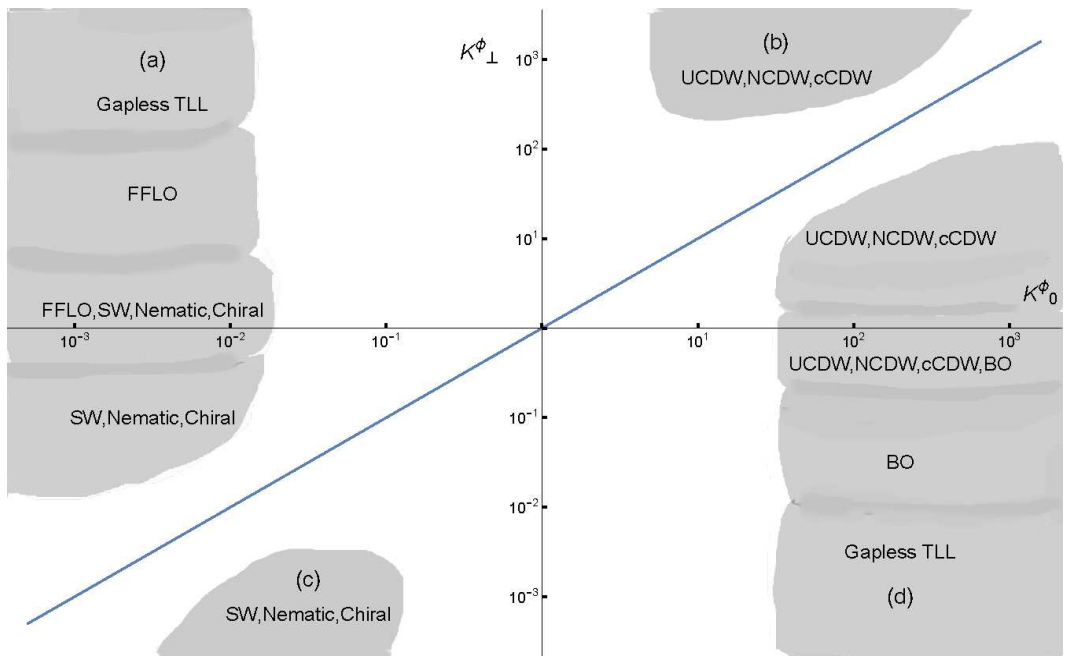}
\par\end{centering}
\caption{\label{fig:pd} The figure shows the phase diagram for a system of
three coupled spinless Luttinger liquids as a function of $K_{\bot}^{\phi}$, considering the parameter regimes
(a) $K_{\perp}^{\phi}\gg K_{0}^{\phi}$ and $K_{0}^{\phi}\ll1$, where
only particle-particle (p-p) orders are considered due to the smaller
scaling dimensions of the corresponding order parameters, (b) $K_{\perp}^{\phi}\gg K_{0}^{\phi}$ and $K_{0}^{\phi}\gg1$, where
the dominant instabilities belong to the intraband
particle-hole channel, (c) $K_{0}^{\phi}\gg K_{\bot}^{\phi}$
and $K_{0}^{\phi}\ll1$ where the dominant instabilities occur in the intraband particle-particle channel, and
 (d) $K_{0}^{\phi}\gg K_{\bot}^{\phi}$
and $K_{0}^{\phi}\gg1$, where only particle-hole (p-h) orders are
considered in our analysis, due to smaller scaling dimensions of the corresponding terms. In cases (a) 
and (d), we can tune $K_{\bot}^{\phi}$ over a large range of values,
and for $K_{\bot}^{\phi}\sim1$, various interband and intraband
orders compete with one another, the winner being determined
by the initial conditions on the interactions. Note that our results are not reliable for $K_{\bot}^{\phi}=1$ in regime (a), where the one-loop corrections must be taken into account. The orders indicated in the figure have 
been denoted in the paper as SW for $s-$wave, FFLO for finite-momentum pairing, UCDW as a CDW order with s-wave symmetry, NCDW as nematic $d-$density wave, cCDW as chiral $d-$density wave and BO as bond order. The  shaded (gray) portions of the phase diagram demarcate the parameter regimes which can be understood from our analysis.The boundaries of different types of phases are flexible in nature, and can change depending on the initial conditions chosen for the couplings. }
\end{figure*}


Below we discuss the salient features of the phase diagram for the aforementioned two parameter regimes, $K_{0}^{\phi}\gg K_{\bot}^{\phi}$ and $K_{0}^{\phi}\gg1$,
or $K_{0}^{\phi}\ll K_{\bot}^{\phi}$ and $K_{0}^{\phi}\ll1$, each corresponding to a range of values of $K_{\bot}^{\phi}$. Since
$K_{\bot}^{\theta}$ is inversely related to $K_{\bot}^{\phi}$ in
our model, it does not constitute an independent parameter in the
phase diagram. 

\paragraph*{$K_{\bot}^{\phi}\ll1$:}

In this regime, for $K_{0}^{\phi}\ll K_{\bot}^{\phi}$ and $K_{0}^{\phi}\ll1$,
the intraband particle-particle orders (SW, Nematic, Chiral) are found to be more relevant, whereas
for $K_{0}^{\phi}\gg K_{\bot}^{\phi}$ and $K_{0}^{\phi}\gg1$, 
no electronic orders are present when we consider extremely small values of $K_{\bot}^{\phi}$,
and for larger values of $K_{\bot}^{\phi}$, a particular pair of
interband particle-hole orders (BO) dominates, depending upon the initial
conditions being considered for the interactions. 

\paragraph*{$K_{\bot}^{\phi}\sim1$: }

For $K_{\bot}^{\phi}\sim1$, various intraband and interband particle-particle (FFLO, SW, Nematic, Chiral) orders compete with each other in the regime $K_{0}^{\phi}\ll K_{\bot}^{\phi}$ and $K_{0}^{\phi}\ll1$, and likewise, various particle-hole (UCDW, NCDW, cCDW,BO) orders compete with each other in the regime  $K_{0}^{\phi}\gg K_{\bot}^{\phi}$ and $K_{0}^{\phi}\gg1$, and it is in this part of the phase diagram that the winning phases
are dependent most sensitively on the initial conditions chosen for
the interactions. However, at $K_{\bot}^{\phi}=1$ for $K_{0}^{\phi}\ll K_{\bot}^{\phi}$, or very close to this point, the one-loop corrections should be taken into account, and our analysis in this regime requires further work. 

\paragraph*{$K_{\bot}^{\phi}\gg1$: }

In this case, for $K_{0}^{\phi}\ll K_{\bot}^{\phi}$ and $K_{0}^{\phi}\ll1$,
a particular pair of interband particle-particle orders (FFLO) is found to
dominate, depending on the initial conditions
chosen for the interactions, and no order is found to be present when we consider extremely
large values of $K_{\bot}^{\phi}$, whereas for $K_{0}^{\phi}\gg K_{\bot}^{\phi}$ and $K_{0}^{\phi}\gg1$,
the intraband particle-hole orders (UCDW, NCDW, cCDW) are found to be more relevant. 

The types of electronic orders occurring in different parameter
regimes, considered in our analysis, are schematically shown in Fig. \ref{fig:pd}. 

\section{\label{sec:discussion-and-conclusions}discussion and conclusions}

In summary, we have studied competing electronic phases and phase
transitions in a system of three coupled spinless Luttinger liquids
using a renormalization group analysis of the bosonized interactions
that takes into account off-diagonal contributions arising from one-loop corrections to the stiffness matrices. This is done by introducing a series of rotations and rescalings of the fields (or equivalently, the
coefficients of different fields in the sine-Gordon interaction terms) in the course of the RG flow.
These rotations and rescalings are found to depend on all the
couplings as well as coefficients of all the fields present in the system. They couple the different interaction channels even at the tree-level order. To determine the most dominant electronic orders, we introduce,
in the disordered phase, test vertices corresponding to
various order parameter fluctuations and study their evolution under the renormalization group. We find that the overall nature
of the winning orders in different parameter regimes is governed by
the RG flows of the couplings, as well as those of the coefficients
of the fields in the sine-Gordon terms. Notably, for a range of values of the Luttinger liquid parameter $K_{\bot}^{\phi}$, which depart appreciably from the noninteracting limit $K_{\bot}^{\phi}=1,$ interband
orders involving any one
pair of bands are found to be dominant, the specific pair being determined by the initial conditions
for the couplings. This is an example of valley symmetry breaking. At $K_{\bot}^{\phi}=1$ for $K_{\perp}^{\phi}\gg K_{0}^{\phi}$, one-loop corrections to the RG equations must be taken into account, and this aspect of our analysis requires further work. In the regions where intraband orders are the most relevant, they can be chiral in nature. Such orders restore the original $C_{3}$ symmetry of the system, broken explicitly through the initial conditions for the couplings. In the regimes where $K_{\bot}^{\phi}\sim1$,
the nature of the dominant orders is found to be sensitively determined by the initial conditions on the interaction couplings, with multiple orders competing closely. For simplicity of analysis, we have considered
the strong correlation regimes of $K_{0}^{\phi}\gg1$ or $K_{0}^{\phi}\ll1$, and the more involved
case of $K_{0}^{\phi}\sim1$ has not been discussed, where the particle-particle and particle-hole channels compete
with each other and the results are likely to be sensitive to the
initial conditions considered. This will be taken up in a future work.
We also determine the nature of the phase transitions as a function of
the Luttinger parameter $K_{\perp}^{\phi}$ as well as the initial
conditions on the interactions $g_{\alpha}$ using a numerical scaling
analysis.  The system hosts continuous transitions belonging to the BKT universality class, where the critical value of $K_{\bot}^{\phi}$ differs with the initial values of the couplings.

From an experimental point of view, our analysis is expected to be relevant for studying electronic interaction effects in semimetals with three small Fermi pockets under conditions of high magnetic fields such that the bands are effectively in the quantum limit, and may be regarded as one-dimensional. Examples include bismuth, the graphite intercalation compounds and possibly the heavy fermion semimetal UTe$_{2}$ at high magnetic fields.  For bismuth, when the magnetic field is aligned along the highest symmetry axis (the trigonal axis), a field of 9 T allows one to attain the
quantum limit putting carriers in their lowest Landau level.\cite{Yang2010} In this situation,
Coulomb interaction effects play an important role in determining the electronic phase.  The presence of anomalous features in the magnetization \cite{Li2008} and the Nernst response \cite{Behnia1729} of bismuth at high fields, beyond the quantum limit, points towards the importance of examining possible electronic instabilities due to interaction effects in this regime. 
Furthermore, there has been experimental evidence for valley symmetry breaking at high magnetic fields in bismuth, \cite{Kuchler2014} and the importance of electron correlations for the same has been recognized.
From recent magnetoresistance studies, one or two valleys have been observed to become completely empty above a threshold magnetic field. \cite{Zhu2018}
Moreover, in semi-metallic bismuth the flow of Dirac fermions along the trigonal axis 
is extremely sensitive to the orientation of in-plane magnetic field. 
In the vicinity of the quantum limit, the orientation of magnetic field significantly affects the distribution of carriers in each valley, 
and the valley polarization is induced by the magnetic field. As the temperature is decreased or the magnetic field increased, the symmetry between the three valleys is spontaneously lost. We expect our technique to be useful for theoretically describing such a situation in bismuth, incorporating the features known from experiment, and predicting possible electronic instabilities. 

In graphite intercalates, the Fermi level often naturally lies in the vicinity of the M-points in the Brillouin zone, which gives rise to another system with three small Fermi pockets. Superconductivity has been predicted and observed experimentally in multiple graphite intercalation compounds, such as CaC$_{6}$,YbC$_{6}$ and KC$_{8}$, \cite{Weller2005} but the possibility of realizing superconductivity or a density wave order under a high magnetic field in such materials has not received much attention in the literature. The case of pure graphite is different; there is evidence for a high field-induced CDW transition \cite{Yoshioka1981} resulting from the enhancement of interactions due to the confinement effect of the magnetic field. However valley-symmetry breaking in graphite occurs between the K and K$^{\prime}$  points, which is not the subject of this paper.
Corresponding field-induced phase transitions in graphite intercalates may, however, be accessible using our analysis. 

The recently discovered heavy fermion triplet superconductor UTe$_{2}$,\cite{Ran684, Aoki2019, Metz2019,Ishizuka2019,Jiao2019, Sundar2019}  with a transition temperature $T_{sc}$=1.6 K, exhibits two independent high-field superconducting phases,\cite{Ran2019} one of which has an upper critical field exceeding 65 T, and lies within a field-polarized phase. Such re-entrant superconducting phases are observed for selective ranges of orientation of the field. \cite{Ran2019, Knebel2019} High-resolution ARPES data for UTe$_{2}$ indicates that it has three small Fermi pockets.\cite{Miao2020} A quasi-1D bandstructure has been indicated both by bandstructure calculations and the ARPES studies. Our analysis is expected to be applicable at the highest fields, with electrons fully spin-polarized and in the quantum limit. A field of 65 T corresponds to a magnetic length of about 3.2 nm, which would require a carrier density of about 7x10$^{18}$ cm$^{-3}$ to be in the quantum limit, typical for semimetallic systems\cite{Akiba2015}. 

In the present work, we have not studied the case where $K_{0}^{\phi}\sim K_{\bot}^{\phi}$,
with the rotations being in general O(3) matrices. The rotation matrices
in that case are non-abelian and it would interesting to see if this
gives qualitatively new insights into the problem. In this regime,
we also have the possibility of an additional Ising-type symmetry
breaking due to the symmetry between the $\widetilde{\theta}$ and
$\widetilde{\phi}$ fields when $K_{\bot}^{\phi}=K_{0}^{\phi}=1$,
which has not been considered in this paper. We hope to study the implications of our approach for
the spinful three-band case, and compare our results with Ref. \onlinecite{PhysRevB.94.205129},
where the rotations of the matrices $Z_{\mu\nu}$ were not taken into
account in the RG analysis. We would also like to consider the case
of special fillings where intraband Umklapp scattering terms are possible.
At first sight, these terms have higher scaling dimensions than the
interactions considered by us, and so, at the tree level, they are
not relevant. However, more work needs to be done to see the effect
they have on the conclusions of this paper. 
\begin{acknowledgments}
VT acknowledges DST for a Swarnajayanti grant (No. DST/SJF/PSA-0212012-13). 
\end{acknowledgments}

\bibliographystyle{apsrev4-1}

\begin{thebibliography}{45}%
\makeatletter
\providecommand \@ifxundefined [1]{%
 \@ifx{#1\undefined}
}%
\providecommand \@ifnum [1]{%
 \ifnum #1\expandafter \@firstoftwo
 \else \expandafter \@secondoftwo
 \fi
}%
\providecommand \@ifx [1]{%
 \ifx #1\expandafter \@firstoftwo
 \else \expandafter \@secondoftwo
 \fi
}%
\providecommand \natexlab [1]{#1}%
\providecommand \enquote  [1]{``#1''}%
\providecommand \bibnamefont  [1]{#1}%
\providecommand \bibfnamefont [1]{#1}%
\providecommand \citenamefont [1]{#1}%
\providecommand \href@noop [0]{\@secondoftwo}%
\providecommand \href [0]{\begingroup \@sanitize@url \@href}%
\providecommand \@href[1]{\@@startlink{#1}\@@href}%
\providecommand \@@href[1]{\endgroup#1\@@endlink}%
\providecommand \@sanitize@url [0]{\catcode `\\12\catcode `\$12\catcode
  `\&12\catcode `\#12\catcode `\^12\catcode `\_12\catcode `\%12\relax}%
\providecommand \@@startlink[1]{}%
\providecommand \@@endlink[0]{}%
\providecommand \url  [0]{\begingroup\@sanitize@url \@url }%
\providecommand \@url [1]{\endgroup\@href {#1}{\urlprefix }}%
\providecommand \urlprefix  [0]{URL }%
\providecommand \Eprint [0]{\href }%
\providecommand \doibase [0]{http://dx.doi.org/}%
\providecommand \selectlanguage [0]{\@gobble}%
\providecommand \bibinfo  [0]{\@secondoftwo}%
\providecommand \bibfield  [0]{\@secondoftwo}%
\providecommand \translation [1]{[#1]}%
\providecommand \BibitemOpen [0]{}%
\providecommand \bibitemStop [0]{}%
\providecommand \bibitemNoStop [0]{.\EOS\space}%
\providecommand \EOS [0]{\spacefactor3000\relax}%
\providecommand \BibitemShut  [1]{\csname bibitem#1\endcsname}%
\let\auto@bib@innerbib\@empty
\bibitem [{\citenamefont {Nishimoto}\ \emph {et~al.}(2009)\citenamefont
  {Nishimoto}, \citenamefont {Jeckelmann},\ and\ \citenamefont
  {J.~Scalapino}}]{article}%
  \BibitemOpen
  \bibfield  {author} {\bibinfo {author} {\bibfnamefont {S.}~\bibnamefont
  {Nishimoto}}, \bibinfo {author} {\bibfnamefont {E.}~\bibnamefont
  {Jeckelmann}}, \ and\ \bibinfo {author} {\bibfnamefont {D.}~\bibnamefont
  {J.~Scalapino}},\ }\href@noop {} {\bibfield  {journal} {\bibinfo  {journal}
  {Phys. Rev. B}\ }\textbf {\bibinfo {volume} {79}} (\bibinfo {year}
  {2009})}\BibitemShut {NoStop}%
\bibitem [{\citenamefont {Fjaerestad}\ and\ \citenamefont
  {Marston}(2002)}]{PhysRevB.65.125106}%
  \BibitemOpen
  \bibfield  {author} {\bibinfo {author} {\bibfnamefont {J.~O.}\ \bibnamefont
  {Fjaerestad}}\ and\ \bibinfo {author} {\bibfnamefont {J.~B.}\ \bibnamefont
  {Marston}},\ }\href@noop {} {\bibfield  {journal} {\bibinfo  {journal} {Phys.
  Rev. B}\ }\textbf {\bibinfo {volume} {65}},\ \bibinfo {pages} {125106}
  (\bibinfo {year} {2002})}\BibitemShut {NoStop}%
\bibitem [{\citenamefont {Nishimoto}\ \emph {et~al.}(2002)\citenamefont
  {Nishimoto}, \citenamefont {Jeckelmann},\ and\ \citenamefont
  {Scalapino}}]{PhysRevB.66.245109}%
  \BibitemOpen
  \bibfield  {author} {\bibinfo {author} {\bibfnamefont {S.}~\bibnamefont
  {Nishimoto}}, \bibinfo {author} {\bibfnamefont {E.}~\bibnamefont
  {Jeckelmann}}, \ and\ \bibinfo {author} {\bibfnamefont {D.~J.}\ \bibnamefont
  {Scalapino}},\ }\href@noop {} {\bibfield  {journal} {\bibinfo  {journal}
  {Phys. Rev. B}\ }\textbf {\bibinfo {volume} {66}},\ \bibinfo {pages} {245109}
  (\bibinfo {year} {2002})}\BibitemShut {NoStop}%
\bibitem [{\citenamefont {Lee}\ \emph {et~al.}(2005)\citenamefont {Lee},
  \citenamefont {Marston},\ and\ \citenamefont
  {Fjaerestad}}]{PhysRevB.72.075126}%
  \BibitemOpen
  \bibfield  {author} {\bibinfo {author} {\bibfnamefont {S.}~\bibnamefont
  {Lee}}, \bibinfo {author} {\bibfnamefont {J.~B.}\ \bibnamefont {Marston}}, \
  and\ \bibinfo {author} {\bibfnamefont {J.~O.}\ \bibnamefont {Fjaerestad}},\
  }\href@noop {} {\bibfield  {journal} {\bibinfo  {journal} {Phys. Rev. B}\
  }\textbf {\bibinfo {volume} {72}},\ \bibinfo {pages} {075126} (\bibinfo
  {year} {2005})}\BibitemShut {NoStop}%
\bibitem [{\citenamefont {Chudzinski}\ \emph {et~al.}(2007)\citenamefont
  {Chudzinski}, \citenamefont {Gabay},\ and\ \citenamefont
  {Giamarchi}}]{PhysRevB.76.161101}%
  \BibitemOpen
  \bibfield  {author} {\bibinfo {author} {\bibfnamefont {P.}~\bibnamefont
  {Chudzinski}}, \bibinfo {author} {\bibfnamefont {M.}~\bibnamefont {Gabay}}, \
  and\ \bibinfo {author} {\bibfnamefont {T.}~\bibnamefont {Giamarchi}},\
  }\href@noop {} {\bibfield  {journal} {\bibinfo  {journal} {Phys. Rev. B}\
  }\textbf {\bibinfo {volume} {76}},\ \bibinfo {pages} {161101} (\bibinfo
  {year} {2007})}\BibitemShut {NoStop}%
\bibitem [{\citenamefont {Chudzinski}\ \emph {et~al.}(2008)\citenamefont
  {Chudzinski}, \citenamefont {Gabay},\ and\ \citenamefont
  {Giamarchi}}]{PhysRevB.78.075124}%
  \BibitemOpen
  \bibfield  {author} {\bibinfo {author} {\bibfnamefont {P.}~\bibnamefont
  {Chudzinski}}, \bibinfo {author} {\bibfnamefont {M.}~\bibnamefont {Gabay}}, \
  and\ \bibinfo {author} {\bibfnamefont {T.}~\bibnamefont {Giamarchi}},\
  }\href@noop {} {\bibfield  {journal} {\bibinfo  {journal} {Phys. Rev. B}\
  }\textbf {\bibinfo {volume} {78}},\ \bibinfo {pages} {075124} (\bibinfo
  {year} {2008})}\BibitemShut {NoStop}%
  \bibitem [{\citenamefont {Assaraf}\ \emph {et~al.}(1999)\citenamefont
  {Assaraf}, \citenamefont {Azaria},\citenamefont {Caffarel},\ and\ \citenamefont
  {Lecheminant}}]{PhysRevB.60.2299}%
  \BibitemOpen
  \bibfield  {author} {\bibinfo {author} {\bibfnamefont {R.}~\bibnamefont
  {Assaraf}}, \bibinfo {author} {\bibfnamefont {P.}~\bibnamefont {Azaria}}, \bibinfo {author} {\bibfnamefont {M.}~\bibnamefont {Caffarel}}, \
  and\ \bibinfo {author} {\bibfnamefont {P.}~\bibnamefont {Lecheminant}},\
  }\href@noop {} {\bibfield  {journal} {\bibinfo  {journal} {Phys. Rev. B}\
  }\textbf {\bibinfo {volume} {60}},\ \bibinfo {pages} {2299} (\bibinfo
  {year} {1999})}\BibitemShut {NoStop}%
\bibitem [{\citenamefont {Tsuchiizu}\ and\ \citenamefont
  {Furusaki}(2002)}]{PhysRevB.66.245106}%
  \BibitemOpen
  \bibfield  {author} {\bibinfo {author} {\bibfnamefont {M.}~\bibnamefont
  {Tsuchiizu}}\ and\ \bibinfo {author} {\bibfnamefont {A.}~\bibnamefont
  {Furusaki}},\ }\href@noop {} {\bibfield  {journal} {\bibinfo  {journal}
  {Phys. Rev. B}\ }\textbf {\bibinfo {volume} {66}},\ \bibinfo {pages} {245106}
  (\bibinfo {year} {2002})}\BibitemShut {NoStop}%
\bibitem [{\citenamefont {Wessel}\ \emph {et~al.}(2003)\citenamefont {Wessel},
  \citenamefont {Indergand}, \citenamefont {Lauchli}, \citenamefont
  {Ledermann},\ and\ \citenamefont {Sigrist}}]{PhysRevB.67.184517}%
  \BibitemOpen
  \bibfield  {author} {\bibinfo {author} {\bibfnamefont {S.}~\bibnamefont
  {Wessel}}, \bibinfo {author} {\bibfnamefont {M.}~\bibnamefont {Indergand}},
  \bibinfo {author} {\bibfnamefont {A.}~\bibnamefont {Lauchli}}, \bibinfo
  {author} {\bibfnamefont {U.}~\bibnamefont {Ledermann}}, \ and\ \bibinfo
  {author} {\bibfnamefont {M.}~\bibnamefont {Sigrist}},\ }\href@noop {}
  {\bibfield  {journal} {\bibinfo  {journal} {Phys. Rev. B}\ }\textbf {\bibinfo
  {volume} {67}},\ \bibinfo {pages} {184517} (\bibinfo {year}
  {2003})}\BibitemShut {NoStop}%
\bibitem [{\citenamefont {Wu}\ \emph {et~al.}(2003)\citenamefont {Wu},
  \citenamefont {Vincent~Liu},\ and\ \citenamefont
  {Fradkin}}]{PhysRevB.68.115104}%
  \BibitemOpen
  \bibfield  {author} {\bibinfo {author} {\bibfnamefont {C.}~\bibnamefont
  {Wu}}, \bibinfo {author} {\bibfnamefont {W.}~\bibnamefont {Vincent~Liu}}, \
  and\ \bibinfo {author} {\bibfnamefont {E.}~\bibnamefont {Fradkin}},\
  }\href@noop {} {\bibfield  {journal} {\bibinfo  {journal} {Phys. Rev. B}\
  }\textbf {\bibinfo {volume} {68}},\ \bibinfo {pages} {115104} (\bibinfo
  {year} {2003})}\BibitemShut {NoStop}%
\bibitem [{\citenamefont {O'Hern}\ \emph {et~al.}(1999)\citenamefont {O'Hern},
  \citenamefont {Lubensky},\ and\ \citenamefont {Toner}}]{PhysRevLett.83.2745}%
  \BibitemOpen
  \bibfield  {author} {\bibinfo {author} {\bibfnamefont {C.~S.}\ \bibnamefont
  {O'Hern}}, \bibinfo {author} {\bibfnamefont {T.~C.}\ \bibnamefont
  {Lubensky}}, \ and\ \bibinfo {author} {\bibfnamefont {J.}~\bibnamefont
  {Toner}},\ }\href@noop {} {\bibfield  {journal} {\bibinfo  {journal} {Phys.
  Rev. Lett.}\ }\textbf {\bibinfo {volume} {83}},\ \bibinfo {pages} {2745}
  (\bibinfo {year} {1999})}\BibitemShut {NoStop}%
\bibitem [{\citenamefont {Carpentier}\ and\ \citenamefont
  {Orignac}(2006)}]{PhysRevB.74.085409}%
  \BibitemOpen
  \bibfield  {author} {\bibinfo {author} {\bibfnamefont {D.}~\bibnamefont
  {Carpentier}}\ and\ \bibinfo {author} {\bibfnamefont {E.}~\bibnamefont
  {Orignac}},\ }\href@noop {} {\bibfield  {journal} {\bibinfo  {journal} {Phys.
  Rev. B}\ }\textbf {\bibinfo {volume} {74}},\ \bibinfo {pages} {085409}
  (\bibinfo {year} {2006})}\BibitemShut {NoStop}%
\bibitem [{\citenamefont {DeGottardi}\ \emph {et~al.}(2010)\citenamefont
  {DeGottardi}, \citenamefont {Wei}, \citenamefont {Fernandez},\ and\
  \citenamefont {Vishveshwara}}]{PhysRevB.82.155411}%
  \BibitemOpen
  \bibfield  {author} {\bibinfo {author} {\bibfnamefont {W.}~\bibnamefont
  {DeGottardi}}, \bibinfo {author} {\bibfnamefont {T.-C.}\ \bibnamefont {Wei}},
  \bibinfo {author} {\bibfnamefont {V.}~\bibnamefont {Fernandez}}, \ and\
  \bibinfo {author} {\bibfnamefont {S.}~\bibnamefont {Vishveshwara}},\
  }\href@noop {} {\bibfield  {journal} {\bibinfo  {journal} {Phys. Rev. B}\
  }\textbf {\bibinfo {volume} {82}},\ \bibinfo {pages} {155411} (\bibinfo
  {year} {2010})}\BibitemShut {NoStop}%
\bibitem [{\citenamefont {Suzumura}\ and\ \citenamefont
  {Tsuchiizu}(2001)}]{SUZUMURA200193}%
  \BibitemOpen
  \bibfield  {author} {\bibinfo {author} {\bibfnamefont {Y.}~\bibnamefont
  {Suzumura}}\ and\ \bibinfo {author} {\bibfnamefont {M.}~\bibnamefont
  {Tsuchiizu}},\ }\href@noop {} {\bibfield  {journal} {\bibinfo  {journal}
  {Journal of Physics and Chemistry of Solids}\ }\textbf {\bibinfo {volume}
  {62}},\ \bibinfo {pages} {93 } (\bibinfo {year} {2001})}\BibitemShut
  {NoStop}%
\bibitem [{\citenamefont {Chen}\ \emph {et~al.}(2001)\citenamefont {Chen},
  \citenamefont {Buttner},\ and\ \citenamefont {Voit}}]{PhysRevLett.87.087205}%
  \BibitemOpen
  \bibfield  {author} {\bibinfo {author} {\bibfnamefont {S.}~\bibnamefont
  {Chen}}, \bibinfo {author} {\bibfnamefont {H.}~\bibnamefont {Buttner}}, \
  and\ \bibinfo {author} {\bibfnamefont {J.}~\bibnamefont {Voit}},\ }\href@noop
  {} {\bibfield  {journal} {\bibinfo  {journal} {Phys. Rev. Lett.}\ }\textbf
  {\bibinfo {volume} {87}},\ \bibinfo {pages} {087205} (\bibinfo {year}
  {2001})}\BibitemShut {NoStop}%
\bibitem [{\citenamefont {Vincent~Liu}\ and\ \citenamefont
  {Fradkin}(2001)}]{PhysRevLett.86.1865}%
  \BibitemOpen
  \bibfield  {author} {\bibinfo {author} {\bibfnamefont {W.}~\bibnamefont
  {Vincent~Liu}}\ and\ \bibinfo {author} {\bibfnamefont {E.}~\bibnamefont
  {Fradkin}},\ }\href@noop {} {\bibfield  {journal} {\bibinfo  {journal} {Phys.
  Rev. Lett.}\ }\textbf {\bibinfo {volume} {86}},\ \bibinfo {pages} {1865}
  (\bibinfo {year} {2001})}\BibitemShut {NoStop}%
\bibitem [{\citenamefont {Sheng}\ \emph {et~al.}(2009)\citenamefont {Sheng},
  \citenamefont {Motrunich},\ and\ \citenamefont
  {Fisher}}]{PhysRevB.79.205112}%
  \BibitemOpen
  \bibfield  {author} {\bibinfo {author} {\bibfnamefont {D.~N.}\ \bibnamefont
  {Sheng}}, \bibinfo {author} {\bibfnamefont {O.~I.}\ \bibnamefont
  {Motrunich}}, \ and\ \bibinfo {author} {\bibfnamefont {M.~P.~A.}\
  \bibnamefont {Fisher}},\ }\href@noop {} {\bibfield  {journal} {\bibinfo
  {journal} {Phys. Rev. B}\ }\textbf {\bibinfo {volume} {79}},\ \bibinfo
  {pages} {205112} (\bibinfo {year} {2009})}\BibitemShut {NoStop}%
\bibitem [{\citenamefont {Sato}(2007)}]{PhysRevB.76.054427}%
  \BibitemOpen
  \bibfield  {author} {\bibinfo {author} {\bibfnamefont {M.}~\bibnamefont
  {Sato}},\ }\href@noop {} {\bibfield  {journal} {\bibinfo  {journal} {Phys.
  Rev. B}\ }\textbf {\bibinfo {volume} {76}},\ \bibinfo {pages} {054427}
  (\bibinfo {year} {2007})}\BibitemShut {NoStop}%
\bibitem [{\citenamefont {Shelton}\ \emph {et~al.}(1996)\citenamefont
  {Shelton}, \citenamefont {Nersesyan},\ and\ \citenamefont
  {Tsvelik}}]{PhysRevB.53.8521}%
  \BibitemOpen
  \bibfield  {author} {\bibinfo {author} {\bibfnamefont {D.~G.}\ \bibnamefont
  {Shelton}}, \bibinfo {author} {\bibfnamefont {A.~A.}\ \bibnamefont
  {Nersesyan}}, \ and\ \bibinfo {author} {\bibfnamefont {A.~M.}\ \bibnamefont
  {Tsvelik}},\ }\href@noop {} {\bibfield  {journal} {\bibinfo  {journal} {Phys.
  Rev. B}\ }\textbf {\bibinfo {volume} {53}},\ \bibinfo {pages} {8521}
  (\bibinfo {year} {1996})}\BibitemShut {NoStop}%
\bibitem [{\citenamefont {Khveshchenko}\ and\ \citenamefont
  {Rice}(1994)}]{PhysRevB.50.252}%
  \BibitemOpen
  \bibfield  {author} {\bibinfo {author} {\bibfnamefont {D.~V.}\ \bibnamefont
  {Khveshchenko}}\ and\ \bibinfo {author} {\bibfnamefont {T.~M.}\ \bibnamefont
  {Rice}},\ }\href@noop {} {\bibfield  {journal} {\bibinfo  {journal} {Phys.
  Rev. B}\ }\textbf {\bibinfo {volume} {50}},\ \bibinfo {pages} {252} (\bibinfo
  {year} {1994})}\BibitemShut {NoStop}%
\bibitem [{\citenamefont {Cabra}\ \emph {et~al.}(2000)\citenamefont {Cabra},
  \citenamefont {Honecker},\ and\ \citenamefont {Pujol}}]{Cabra2000}%
  \BibitemOpen
  \bibfield  {author} {\bibinfo {author} {\bibfnamefont {D.}~\bibnamefont
  {Cabra}}, \bibinfo {author} {\bibfnamefont {A.}~\bibnamefont {Honecker}}, \
  and\ \bibinfo {author} {\bibfnamefont {P.}~\bibnamefont {Pujol}},\
  }\href@noop {} {\bibfield  {journal} {\bibinfo  {journal} {The European
  Physical Journal B - Condensed Matter and Complex Systems}\ }\textbf
  {\bibinfo {volume} {13}},\ \bibinfo {pages} {55} (\bibinfo {year}
  {2000})}\BibitemShut {NoStop}%
\bibitem [{\citenamefont {Allen}\ \emph {et~al.}(2001)\citenamefont {Allen},
  \citenamefont {Azaria},\ and\ \citenamefont {Lecheminant}}]{Allen_2001}%
  \BibitemOpen
  \bibfield  {author} {\bibinfo {author} {\bibfnamefont {D.}~\bibnamefont
  {Allen}}, \bibinfo {author} {\bibfnamefont {P.}~\bibnamefont {Azaria}}, \
  and\ \bibinfo {author} {\bibfnamefont {P.}~\bibnamefont {Lecheminant}},\
  }\href@noop {} {\bibfield  {journal} {\bibinfo  {journal} {Journal of Physics
  A: Mathematical and General}\ }\textbf {\bibinfo {volume} {34}},\ \bibinfo
  {pages} {L305} (\bibinfo {year} {2001})}\BibitemShut {NoStop}%
\bibitem [{\citenamefont {Miao}\ \emph {et~al.}(2016)\citenamefont {Miao},
  \citenamefont {Zhang},\ and\ \citenamefont {Zhou}}]{PhysRevB.94.205129}%
  \BibitemOpen
  \bibfield  {author} {\bibinfo {author} {\bibfnamefont {J.-J.}\ \bibnamefont
  {Miao}}, \bibinfo {author} {\bibfnamefont {F.-C.}\ \bibnamefont {Zhang}}, \
  and\ \bibinfo {author} {\bibfnamefont {Y.}~\bibnamefont {Zhou}},\ }\href@noop
  {} {\bibfield  {journal} {\bibinfo  {journal} {Phys. Rev. B}\ }\textbf
  {\bibinfo {volume} {94}},\ \bibinfo {pages} {205129} (\bibinfo {year}
  {2016})}\BibitemShut {NoStop}%
\bibitem [{\citenamefont {Le~Hur}\ \emph {et~al.}(2017)\citenamefont {Le~Hur},
  \citenamefont {Soret},\ and\ \citenamefont {Yang}}]{PhysRevB.96.205109}%
  \BibitemOpen
  \bibfield  {author} {\bibinfo {author} {\bibfnamefont {K.}~\bibnamefont
  {Le~Hur}}, \bibinfo {author} {\bibfnamefont {A.}~\bibnamefont {Soret}}, \
  and\ \bibinfo {author} {\bibfnamefont {F.}~\bibnamefont {Yang}},\ }\href@noop
  {} {\bibfield  {journal} {\bibinfo  {journal} {Phys. Rev. B}\ }\textbf
  {\bibinfo {volume} {96}},\ \bibinfo {pages} {205109} (\bibinfo {year}
  {2017})}\BibitemShut {NoStop}%
\bibitem [{\citenamefont {Okamoto}\ and\ \citenamefont
  {Millis}(2012)}]{PhysRevB.85.115406}%
  \BibitemOpen
  \bibfield  {author} {\bibinfo {author} {\bibfnamefont {J.-i.}\ \bibnamefont
  {Okamoto}}\ and\ \bibinfo {author} {\bibfnamefont {A.~J.}\ \bibnamefont
  {Millis}},\ }\href@noop {} {\bibfield  {journal} {\bibinfo  {journal} {Phys.
  Rev. B}\ }\textbf {\bibinfo {volume} {85}},\ \bibinfo {pages} {115406}
  (\bibinfo {year} {2012})}\BibitemShut {NoStop}%
\bibitem [{\citenamefont {Kallin}\ and\ \citenamefont
  {Berlinsky}(2016)}]{Kallin_2016}%
  \BibitemOpen
  \bibfield  {author} {\bibinfo {author} {\bibfnamefont {C.}~\bibnamefont
  {Kallin}}\ and\ \bibinfo {author} {\bibfnamefont {J.}~\bibnamefont
  {Berlinsky}},\ }\href@noop {} {\bibfield  {journal} {\bibinfo  {journal}
  {Reports on Progress in Physics}\ }\textbf {\bibinfo {volume} {79}},\
  \bibinfo {pages} {054502} (\bibinfo {year} {2016})}\BibitemShut {NoStop}%
  \bibitem [{\citenamefont {Arrigoni}(1996)}]{Arrigoni1996}%
  \BibitemOpen
  \bibfield  {author} {\bibinfo {author} {\bibfnamefont {E.}\ \bibnamefont
  {Arrigoni}},\ }\href@noop {} {\bibfield  {journal} {\bibinfo  {journal} {Physica Status Solidi (B)}\ }\textbf {\bibinfo {volume} {195}},\ \bibinfo {pages} {425}
  (\bibinfo {year} {1996})}\BibitemShut {NoStop}%
  \bibitem [{\citenamefont {Arrigoni}(1996)}]{Arrigoni199691}%
  \BibitemOpen
  \bibfield  {author} {\bibinfo {author} {\bibfnamefont {E.}\ \bibnamefont
  {Arrigoni}},\ }\href@noop {} {\bibfield  {journal} {\bibinfo  {journal} {Phys. Lett. A}\ }\textbf {\bibinfo {volume} {215}},\ \bibinfo {pages} {91}
  (\bibinfo {year} {1996})}\BibitemShut {NoStop}%
  \bibitem [{\citenamefont {Kimura}\ \emph {et~al.}(1996)\citenamefont {Kimura},
  \citenamefont {Kuroki},\ and\ \citenamefont {Aoki}}]{Kimura1996}%
  \BibitemOpen
  \bibfield  {author} {\bibinfo {author} {\bibfnamefont {T.}~\bibnamefont
  {Kimura}}, \bibinfo {author} {\bibfnamefont {K.}~\bibnamefont {Kuroki}}, \
  and\ \bibinfo {author} {\bibfnamefont {H.}~\bibnamefont {Aoki}},\ }\href@noop
  {} {\bibfield  {journal} {\bibinfo  {journal} {Phys. Rev. B}\ }\textbf
  {\bibinfo {volume} {54}},\ \bibinfo {pages} {R9608(R)} (\bibinfo {year}
  {1996})}\BibitemShut {NoStop}%
  \bibitem [{\citenamefont {Kimura}\ \emph {et~al.}(1998)\citenamefont {Kimura},
  \citenamefont {Kuroki},\ and\ \citenamefont {Aoki}}]{Kimura1998}%
  \BibitemOpen
  \bibfield  {author} {\bibinfo {author} {\bibfnamefont {T.}~\bibnamefont
  {Kimura}}, \bibinfo {author} {\bibfnamefont {K.}~\bibnamefont {Kuroki}}, \
  and\ \bibinfo {author} {\bibfnamefont {H.}~\bibnamefont {Aoki}},\ }\href@noop
  {} {\bibfield  {journal} {\bibinfo  {journal} {J. Phys. Soc.}\ }\textbf
  {\bibinfo {volume} {67}},\ \bibinfo {pages} {1377} (\bibinfo {year}
  {1998})}\BibitemShut {NoStop}%
  \bibitem [{\citenamefont {Sato}(2007)}]{Sato2007}%
  \BibitemOpen
  \bibfield  {author} {\bibinfo {author} {\bibfnamefont {M.}\ \bibnamefont
  {Sato}},\ }\href@noop {} {\bibfield  {journal} {\bibinfo  {journal} {Phys. Rev. B}\ }\textbf {\bibinfo {volume} {75}},\ \bibinfo {pages} {174407}
  (\bibinfo {year} {2007})}\BibitemShut {NoStop}%
  \bibitem [{\citenamefont {Charrier}\ \emph {et~al.}(2010)\citenamefont {Charrier},
  \citenamefont {Capponi}, \citenamefont {Oshikawa}, \ and\ \citenamefont {Pujol}}]{Charrier2010}%
  \BibitemOpen
  \bibfield  {author} {\bibinfo {author} {\bibfnamefont {D.}~\bibnamefont
  {Charrier}}, \bibinfo {author} {\bibfnamefont {S.}~\bibnamefont {Capponi}},\bibinfo {author} {\bibfnamefont {M.}~\bibnamefont {Oshikawa}}, \
  and\ \bibinfo {author} {\bibfnamefont {P.}~\bibnamefont {Pujol}},\ }\href@noop
  {} {\bibfield  {journal} {\bibinfo  {journal} {Phys. Rev. B}\ }\textbf
  {\bibinfo {volume} {82}},\ \bibinfo {pages} {075108} (\bibinfo {year}
  {2010})}\BibitemShut {NoStop}%
  \bibitem [{\citenamefont {Zhao}\ \emph {et~al.}(2012)\citenamefont {Zhao},
  \citenamefont {Gong}, \citenamefont {Wang}, \ and\ \citenamefont {Su}}]{Zhao2012}%
  \BibitemOpen
  \bibfield  {author} {\bibinfo {author} {\bibfnamefont {Y.}~\bibnamefont
  {Zhao}}, \bibinfo {author} {\bibfnamefont {S-S.}~\bibnamefont {Gong}},\bibinfo {author} {\bibfnamefont {Y-J.}~\bibnamefont {Wang}}, \
  and\ \bibinfo {author} {\bibfnamefont {G.}~\bibnamefont {Su}},\ }\href@noop
  {} {\bibfield  {journal} {\bibinfo  {journal} {Phys. Rev. B}\ }\textbf
  {\bibinfo {volume} {86}},\ \bibinfo {pages} {224406} (\bibinfo {year}
  {2012})}\BibitemShut {NoStop}%
  \bibitem [{\citenamefont {Fuji}\ \emph {et~al.}(2014)\citenamefont {Fuji},
  \citenamefont {Nishimoto}, \citenamefont {Nakada}, \ and\ \citenamefont {Oshikawa}}]{Fuji2014}%
  \BibitemOpen
  \bibfield  {author} {\bibinfo {author} {\bibfnamefont {Y.}~\bibnamefont
  {Fuji}}, \bibinfo {author} {\bibfnamefont {S.}~\bibnamefont {Nishimoto}},\bibinfo {author} {\bibfnamefont {H.}~\bibnamefont {Nakada}}, \
  and\ \bibinfo {author} {\bibfnamefont {M.}~\bibnamefont {Oshikawa}},\ }\href@noop
  {} {\bibfield  {journal} {\bibinfo  {journal} {Phys. Rev. B}\ }\textbf
  {\bibinfo {volume} {89}},\ \bibinfo {pages} {054425} (\bibinfo {year}
  {2014})}\BibitemShut {NoStop}%
  \bibitem [{\citenamefont {Tsukamoto}\ \emph {et~al.}(2000)\citenamefont {Tsukamoto},
  \citenamefont {Kawakami}, \citenamefont {Yamashita}, \ and\ \citenamefont {Ueda}}]{Tsukamoto2000}%
  \BibitemOpen
  \bibfield  {author} {\bibinfo {author} {\bibfnamefont {Y.}~\bibnamefont
  {Tsukamoto}}, \bibinfo {author} {\bibfnamefont {N.}~\bibnamefont {Kawakami}},\bibinfo {author} {\bibfnamefont {Y.}~\bibnamefont {Yamashita}}, \
  and\ \bibinfo {author} {\bibfnamefont {K.}~\bibnamefont {Ueda}},\ }\href@noop
  {} {\bibfield  {journal} {\bibinfo  {journal} {Physica B}\ }\textbf
  {\bibinfo {volume} {281,282}},\ \bibinfo {pages} {540} (\bibinfo {year}
  {2000})}\BibitemShut {NoStop}%
   \bibitem [{\citenamefont {Itoi}\ \emph {et~al.}(1999)\citenamefont {Ioti},
  \citenamefont {Qin},\ and\ \citenamefont {Affleck}}]{Itoi1999}%
  \BibitemOpen
  \bibfield  {author} {\bibinfo {author} {\bibfnamefont {C.}~\bibnamefont
  {Itoi}}, \bibinfo {author} {\bibfnamefont {S.}~\bibnamefont {Qin}}, \
  and\ \bibinfo {author} {\bibfnamefont {I.}~\bibnamefont {Affleck}},\ }\href@noop
  {} {\bibfield  {journal} {\bibinfo  {journal} {Phys.Rev.B}\ }\textbf
  {\bibinfo {volume} {61}},\ \bibinfo {pages} {6747} (\bibinfo {year}
  {1999})}\BibitemShut {NoStop}%
  \bibitem [{\citenamefont {Azaria}\ \emph {et~al.}(1999)\citenamefont {Azaria},
  \citenamefont {Boulat},\ and\ \citenamefont {Lecheminant}}]{Azaria1999}%
  \BibitemOpen
  \bibfield  {author} {\bibinfo {author} {\bibfnamefont {P.}~\bibnamefont
  {Azaria}}, \bibinfo {author} {\bibfnamefont {E.}~\bibnamefont {Boulat}}, \
  and\ \bibinfo {author} {\bibfnamefont {P.}~\bibnamefont {Lecheminant}},\ }\href@noop
  {} {\bibfield  {journal} {\bibinfo  {journal} {Phys. Rev. B}\ }\textbf
  {\bibinfo {volume} {61}},\ \bibinfo {pages} {12112} (\bibinfo {year}
  {1999})}\BibitemShut {NoStop}%
  \bibitem [{\citenamefont {Lee}\ \emph {et~al.}(2004)\citenamefont {Lee},
  \citenamefont {Azaria},\ and\ \citenamefont {Boulat}}]{Lee2004}%
  \BibitemOpen
  \bibfield  {author} {\bibinfo {author} {\bibfnamefont {H.C.}~\bibnamefont
  {Lee}}, \bibinfo {author} {\bibfnamefont {P.}~\bibnamefont {Azaria}}, \
  and\ \bibinfo {author} {\bibfnamefont {E.}~\bibnamefont {Boulat}},\ }\href@noop
  {} {\bibfield  {journal} {\bibinfo  {journal} {Phys. Rev. B}\ }\textbf
  {\bibinfo {volume} {69}},\ \bibinfo {pages} {155109} (\bibinfo {year}
  {2004})}\BibitemShut {NoStop}%
  \bibitem [{\citenamefont {Azaria}\ \emph {et~al.}(1999)\citenamefont {Azaria},
  \citenamefont {Gogolin}, \citenamefont {Lecheminant}, \ and\ \citenamefont {Nersesyan}}]{Azaria19991}%
  \BibitemOpen
  \bibfield  {author} {\bibinfo {author} {\bibfnamefont {P.}~\bibnamefont
  {Azaria}}, \bibinfo {author} {\bibfnamefont {A.O.}~\bibnamefont {Gogolin}},\bibinfo {author} {\bibfnamefont {P.}~\bibnamefont {Lecheminant}}, \
  and\ \bibinfo {author} {\bibfnamefont {A.A.}~\bibnamefont {Nersesyan}},\ }\href@noop
  {} {\bibfield  {journal} {\bibinfo  {journal} {Phys. Rev. Lett.}\ }\textbf
  {\bibinfo {volume} {83}},\ \bibinfo {pages} {624} (\bibinfo {year}
  {1999})}\BibitemShut {NoStop}%
  \bibitem [{\citenamefont {Orignac}\ \emph {et~al.}(1999)\citenamefont {Orignac},
  \citenamefont {Citro},\ and\ \citenamefont {Andrei}}]{Orignac1999}%
  \BibitemOpen
  \bibfield  {author} {\bibinfo {author} {\bibfnamefont {E.}~\bibnamefont
  {Orignac}}, \bibinfo {author} {\bibfnamefont {R.}~\bibnamefont {Citro}}, \
  and\ \bibinfo {author} {\bibfnamefont {N.}~\bibnamefont {Andrei}},\ }\href@noop
  {} {\bibfield  {journal} {\bibinfo  {journal} {Phys. Rev. B}\ }\textbf
  {\bibinfo {volume} {61}},\ \bibinfo {pages} {11533} (\bibinfo {year}
  {1999})}\BibitemShut {NoStop}%
  \bibitem [{\citenamefont {Plat}\ \emph {et~al.}(2015)\citenamefont {Plat},
  \citenamefont {Fuji}, \citenamefont {Capponi}, \ and\ \citenamefont {Pujol}}]{Plat2015}%
  \BibitemOpen
  \bibfield  {author} {\bibinfo {author} {\bibfnamefont {X.}~\bibnamefont
  {Plat}}, \bibinfo {author} {\bibfnamefont {Y.}~\bibnamefont {Fuji}},\bibinfo {author} {\bibfnamefont {S.}~\bibnamefont {Capponi}}, \
  and\ \bibinfo {author} {\bibfnamefont {P.}~\bibnamefont {Pujol}},\ }\href@noop
  {} {\bibfield  {journal} {\bibinfo  {journal} {Phys. Rev. B}\ }\textbf
  {\bibinfo {volume} {91}},\ \bibinfo {pages} {064411} (\bibinfo {year}
  {2015})}\BibitemShut {NoStop}%
  \bibitem [{\citenamefont {Akiba}\ \emph {et~al.}(2015)\citenamefont
  {Akiba}, \citenamefont {Miyake},\citenamefont {Yaguchi},
  \citenamefont {Matsuo},\citenamefont {Kindo},\ and\\citenamefont {Tokunaga}}]{Akiba2015}%
  \BibitemOpen
  \bibfield  {author} {\bibinfo {author} {\bibfnamefont {K.}~\bibnamefont
  {Akiba}}, \bibinfo {author} {\bibfnamefont {A.}~\bibnamefont
  {Miyake}}, \bibinfo {author} {\bibfnamefont {H..}~\bibnamefont
  {Yaguchi}}, \bibinfo {author} {\bibfnamefont {A.}~\bibnamefont
  {Matsuo}}, \bibinfo {author} {\bibfnamefont {K.}~\bibnamefont
  {Kindo}},  \ and\ \bibinfo {author} {\bibfnamefont {M.}~\bibnamefont
  {Tokunaga}},\ }\href@noop {} {\bibfield  {journal} {\bibinfo  {journal}
  {J. Phys. Soc. Jpn.}\ }\textbf {\bibinfo {volume} {84}} (\bibinfo {year}
  {2015})}\BibitemShut {NoStop}%
\bibitem [{\citenamefont {Behnia}\ \emph {et~al.}(2007)\citenamefont {Behnia},
  \citenamefont {Balicas},\ and\ \citenamefont {Kopelevich}}]{Behnia1729}%
  \BibitemOpen
  \bibfield  {author} {\bibinfo {author} {\bibfnamefont {K.}~\bibnamefont
  {Behnia}}, \bibinfo {author} {\bibfnamefont {L.}~\bibnamefont {Balicas}}, \
  and\ \bibinfo {author} {\bibfnamefont {Y.}~\bibnamefont {Kopelevich}},\
  }\href@noop {} {\bibfield  {journal} {\bibinfo  {journal} {Science}\ }\textbf
  {\bibinfo {volume} {317}},\ \bibinfo {pages} {1729} (\bibinfo {year}
  {2007})}\BibitemShut {NoStop}%
\bibitem [{\citenamefont {Fauque}\ \emph
  {et~al.}(2009{\natexlab{a}})\citenamefont {Fauque}, \citenamefont {Vignolle},
  \citenamefont {Proust}, \citenamefont {Issi},\ and\ \citenamefont
  {Behnia}}]{Fauqu__2009}%
  \BibitemOpen
  \bibfield  {author} {\bibinfo {author} {\bibfnamefont {B.}~\bibnamefont
  {Fauque}}, \bibinfo {author} {\bibfnamefont {B.}~\bibnamefont {Vignolle}},
  \bibinfo {author} {\bibfnamefont {C.}~\bibnamefont {Proust}}, \bibinfo
  {author} {\bibfnamefont {J.-P.}\ \bibnamefont {Issi}}, \ and\ \bibinfo
  {author} {\bibfnamefont {K.}~\bibnamefont {Behnia}},\ }\href@noop {}
  {\bibfield  {journal} {\bibinfo  {journal} {New Journal of Physics}\ }\textbf
  {\bibinfo {volume} {11}},\ \bibinfo {pages} {113012} (\bibinfo {year}
  {2009}{\natexlab{a}})}\BibitemShut {NoStop}%
 \bibitem [{\citenamefont {Kuchler}\ \emph {et~al.}(2014)\citenamefont
  {Kuchler}, \citenamefont {Steinke},\citenamefont {Daou},
  \citenamefont {Brando},\citenamefont {Behnia},\ and\\citenamefont {Steglich}}]{Kuchler2014}%
  \BibitemOpen
  \bibfield  {author} {\bibinfo {author} {\bibfnamefont {R.}~\bibnamefont
  {Kuchler}}, \bibinfo {author} {\bibfnamefont {L}~\bibnamefont
  {Steinke}}, \bibinfo {author} {\bibfnamefont {R..}~\bibnamefont
  {Daou}}, \bibinfo {author} {\bibfnamefont {M.}~\bibnamefont
  {Brando}}, \bibinfo {author} {\bibfnamefont {K.}~\bibnamefont
  {Behnia}},  \ and\ \bibinfo {author} {\bibfnamefont {F.}~\bibnamefont
  {Steglich}},\ }\href@noop {} {\bibfield  {journal} {\bibinfo  {journal}
  {Nat. Mater.}\ }\textbf {\bibinfo {volume} {13}} (\bibinfo {year}
  {2014})}\BibitemShut {NoStop}%
\bibitem [{\citenamefont {Li}\ \emph {et~al.}(2008)\citenamefont {Li},
  \citenamefont {Checkelsky}, \citenamefont {Hor}, \citenamefont {Uher},
  \citenamefont {Hebard}, \citenamefont {Cava},\ and\ \citenamefont
  {Ong}}]{Li547}%
  \BibitemOpen
  \bibfield  {author} {\bibinfo {author} {\bibfnamefont {L.}~\bibnamefont
  {Li}}, \bibinfo {author} {\bibfnamefont {J.~G.}\ \bibnamefont {Checkelsky}},
  \bibinfo {author} {\bibfnamefont {Y.~S.}\ \bibnamefont {Hor}}, \bibinfo
  {author} {\bibfnamefont {C.}~\bibnamefont {Uher}}, \bibinfo {author}
  {\bibfnamefont {A.~F.}\ \bibnamefont {Hebard}}, \bibinfo {author}
  {\bibfnamefont {R.~J.}\ \bibnamefont {Cava}}, \ and\ \bibinfo {author}
  {\bibfnamefont {N.~P.}\ \bibnamefont {Ong}},\ }\href@noop {} {\bibfield
  {journal} {\bibinfo  {journal} {Science}\ }\textbf {\bibinfo {volume}
  {321}},\ \bibinfo {pages} {547} (\bibinfo {year} {2008})}\BibitemShut
  {NoStop}%
\bibitem [{\citenamefont {Sharlai}\ and\ \citenamefont
  {Mikitik}(2009)}]{PhysRevB.79.081102}%
  \BibitemOpen
  \bibfield  {author} {\bibinfo {author} {\bibfnamefont {Y.~V.}\ \bibnamefont
  {Sharlai}}\ and\ \bibinfo {author} {\bibfnamefont {G.~P.}\ \bibnamefont
  {Mikitik}},\ }\href@noop {} {\bibfield  {journal} {\bibinfo  {journal} {Phys.
  Rev. B}\ }\textbf {\bibinfo {volume} {79}},\ \bibinfo {pages} {081102}
  (\bibinfo {year} {2009})}\BibitemShut {NoStop}%
\bibitem [{\citenamefont {Alicea}\ and\ \citenamefont
  {Balents}(2009)}]{PhysRevB.79.241101}%
  \BibitemOpen
  \bibfield  {author} {\bibinfo {author} {\bibfnamefont {J.}~\bibnamefont
  {Alicea}}\ and\ \bibinfo {author} {\bibfnamefont {L.}~\bibnamefont
  {Balents}},\ }\href@noop {} {\bibfield  {journal} {\bibinfo  {journal} {Phys.
  Rev. B}\ }\textbf {\bibinfo {volume} {79}},\ \bibinfo {pages} {241101}
  (\bibinfo {year} {2009})}\BibitemShut {NoStop}%
\bibitem [{\citenamefont {Fauque}\ \emph
  {et~al.}(2009{\natexlab{b}})\citenamefont {Fauque}, \citenamefont {Yang},
  \citenamefont {Sheikin}, \citenamefont {Balicas}, \citenamefont {Issi},\ and\
  \citenamefont {Behnia}}]{PhysRevB.79.245124}%
  \BibitemOpen
  \bibfield  {author} {\bibinfo {author} {\bibfnamefont {B.}~\bibnamefont
  {Fauque}}, \bibinfo {author} {\bibfnamefont {H.}~\bibnamefont {Yang}},
  \bibinfo {author} {\bibfnamefont {I.}~\bibnamefont {Sheikin}}, \bibinfo
  {author} {\bibfnamefont {L.}~\bibnamefont {Balicas}}, \bibinfo {author}
  {\bibfnamefont {J.-P.}\ \bibnamefont {Issi}}, \ and\ \bibinfo {author}
  {\bibfnamefont {K.}~\bibnamefont {Behnia}},\ }\href@noop {} {\bibfield
  {journal} {\bibinfo  {journal} {Phys. Rev. B}\ }\textbf {\bibinfo {volume}
  {79}},\ \bibinfo {pages} {245124} (\bibinfo {year}
  {2009}{\natexlab{b}})}\BibitemShut {NoStop}%
\bibitem [{\citenamefont {Matsuo}\ \emph {et~al.}(2009)\citenamefont {Matsuo},
  \citenamefont {Endo}, \citenamefont {Hatano}, \citenamefont {Nakamura},
  \citenamefont {Shirasaki},\ and\ \citenamefont
  {Sugihara}}]{PhysRevB.80.075313}%
  \BibitemOpen
  \bibfield  {author} {\bibinfo {author} {\bibfnamefont {M.}~\bibnamefont
  {Matsuo}}, \bibinfo {author} {\bibfnamefont {A.}~\bibnamefont {Endo}},
  \bibinfo {author} {\bibfnamefont {N.}~\bibnamefont {Hatano}}, \bibinfo
  {author} {\bibfnamefont {H.}~\bibnamefont {Nakamura}}, \bibinfo {author}
  {\bibfnamefont {R.}~\bibnamefont {Shirasaki}}, \ and\ \bibinfo {author}
  {\bibfnamefont {K.}~\bibnamefont {Sugihara}},\ }\href@noop {} {\bibfield
  {journal} {\bibinfo  {journal} {Phys. Rev. B}\ }\textbf {\bibinfo {volume}
  {80}},\ \bibinfo {pages} {075313} (\bibinfo {year} {2009})}\BibitemShut
  {NoStop}%
\bibitem [{\citenamefont {Zhu}\ \emph {et~al.}(2011)\citenamefont {Zhu},
  \citenamefont {Fauque}, \citenamefont {Fuseya},\ and\ \citenamefont
  {Behnia}}]{PhysRevB.84.115137}%
  \BibitemOpen
  \bibfield  {author} {\bibinfo {author} {\bibfnamefont {Z.}~\bibnamefont
  {Zhu}}, \bibinfo {author} {\bibfnamefont {B.}~\bibnamefont {Fauque}},
  \bibinfo {author} {\bibfnamefont {Y.}~\bibnamefont {Fuseya}}, \ and\ \bibinfo
  {author} {\bibfnamefont {K.}~\bibnamefont {Behnia}},\ }\href@noop {}
  {\bibfield  {journal} {\bibinfo  {journal} {Phys. Rev. B}\ }\textbf {\bibinfo
  {volume} {84}},\ \bibinfo {pages} {115137} (\bibinfo {year}
  {2011})}\BibitemShut {NoStop}%
\bibitem [{\citenamefont {Seradjeh}\ \emph {et~al.}(2009)\citenamefont
  {Seradjeh}, \citenamefont {Wu},\ and\ \citenamefont
  {Phillips}}]{PhysRevLett.103.136803}%
  \BibitemOpen
  \bibfield  {author} {\bibinfo {author} {\bibfnamefont {B.}~\bibnamefont
  {Seradjeh}}, \bibinfo {author} {\bibfnamefont {J.}~\bibnamefont {Wu}}, \ and\
  \bibinfo {author} {\bibfnamefont {P.}~\bibnamefont {Phillips}},\ }\href@noop
  {} {\bibfield  {journal} {\bibinfo  {journal} {Phys. Rev. Lett.}\ }\textbf
  {\bibinfo {volume} {103}},\ \bibinfo {pages} {136803} (\bibinfo {year}
  {2009})}\BibitemShut {NoStop}%
\bibitem [{\citenamefont {Yang}\ \emph {et~al.}(2010)\citenamefont {Yang},
  \citenamefont {Fauque}, \citenamefont {Malone}, \citenamefont {Antunes},
  \citenamefont {Zhu}, \citenamefont {Uher},\ and\ \citenamefont
  {Behnia}}]{Yang2010}%
  \BibitemOpen
  \bibfield  {author} {\bibinfo {author} {\bibfnamefont {H.}~\bibnamefont
  {Yang}}, \bibinfo {author} {\bibfnamefont {B.}~\bibnamefont {Fauque}},
  \bibinfo {author} {\bibfnamefont {L.}~\bibnamefont {Malone}}, \bibinfo
  {author} {\bibfnamefont {A.~B.}\ \bibnamefont {Antunes}}, \bibinfo {author}
  {\bibfnamefont {Z.}~\bibnamefont {Zhu}}, \bibinfo {author} {\bibfnamefont
  {C.}~\bibnamefont {Uher}}, \ and\ \bibinfo {author} {\bibfnamefont
  {K.}~\bibnamefont {Behnia}},\ }\href@noop {} {\bibfield  {journal} {\bibinfo
  {journal} {Nature Communications}\ }\textbf {\bibinfo {volume} {1}},\
  \bibinfo {pages} {47} (\bibinfo {year} {2010})}\BibitemShut {NoStop}%
\bibitem [{\citenamefont {Zhu}\ \emph {et~al.}(2012)\citenamefont {Zhu},
  \citenamefont {Fauque}, \citenamefont {Malone}, \citenamefont {Antunes},
  \citenamefont {Fuseya},\ and\ \citenamefont {Behnia}}]{Zhu14813}%
  \BibitemOpen
  \bibfield  {author} {\bibinfo {author} {\bibfnamefont {Z.}~\bibnamefont
  {Zhu}}, \bibinfo {author} {\bibfnamefont {B.}~\bibnamefont {Fauque}},
  \bibinfo {author} {\bibfnamefont {L.}~\bibnamefont {Malone}}, \bibinfo
  {author} {\bibfnamefont {A.~B.}\ \bibnamefont {Antunes}}, \bibinfo {author}
  {\bibfnamefont {Y.}~\bibnamefont {Fuseya}}, \ and\ \bibinfo {author}
  {\bibfnamefont {K.}~\bibnamefont {Behnia}},\ }\href@noop {} {\bibfield
  {journal} {\bibinfo  {journal} {Proceedings of the National Academy of
  Sciences}\ }\textbf {\bibinfo {volume} {109}},\ \bibinfo {pages} {14813}
  (\bibinfo {year} {2012})}\BibitemShut {NoStop}%
\bibitem [{\citenamefont {Dresselhaus}\ and\ \citenamefont
  {Dresselhaus}(1981)}]{dresselhaus1981intercalation}%
  \BibitemOpen
  \bibfield  {author} {\bibinfo {author} {\bibfnamefont {M.}~\bibnamefont
  {Dresselhaus}}\ and\ \bibinfo {author} {\bibfnamefont {G.}~\bibnamefont
  {Dresselhaus}},\ }\href@noop {} {\bibfield  {journal} {\bibinfo  {journal}
  {Advances in Physics}\ }\textbf {\bibinfo {volume} {30}},\ \bibinfo {pages}
  {139} (\bibinfo {year} {1981})}\BibitemShut {NoStop}%
\bibitem [{\citenamefont {Vogel}(1979)}]{Vogel1979}%
  \BibitemOpen
  \bibfield  {author} {\bibinfo {author} {\bibfnamefont {F.~L.}\ \bibnamefont
  {Vogel}},\ }\enquote {\bibinfo {title} {Intercalation compounds of
  graphite},}\ in\ \href@noop {} {\emph {\bibinfo {booktitle} {Molecular
  Metals}}},\ \bibinfo {editor} {edited by\ \bibinfo {editor} {\bibfnamefont
  {W.~E.}\ \bibnamefont {Hatfield}}}\ (\bibinfo  {publisher} {Springer US},\
  \bibinfo {address} {Boston, MA},\ \bibinfo {year} {1979})\ pp.\ \bibinfo
  {pages} {261--279}\BibitemShut {NoStop}%
\bibitem [{\citenamefont {Rao}\ and\ \citenamefont
  {Sen}(2000)}]{rao2000bosonization}%
  \BibitemOpen
  \bibfield  {author} {\bibinfo {author} {\bibfnamefont {S.}~\bibnamefont
  {Rao}}\ and\ \bibinfo {author} {\bibfnamefont {D.}~\bibnamefont {Sen}},\
  }\href@noop {} {\bibfield  {journal} {\bibinfo  {journal}
  {arXiv:cond-mat/0005492}\ } (\bibinfo {year} {2000})}\BibitemShut {NoStop}%
\bibitem [{\citenamefont {von Delft}\ and\ \citenamefont
  {Schoeller}(1998)}]{physik1998}%
  \BibitemOpen
  \bibfield  {author} {\bibinfo {author} {\bibfnamefont {J.}~\bibnamefont {von
  Delft}}\ and\ \bibinfo {author} {\bibfnamefont {H.}~\bibnamefont
  {Schoeller}},\ }\href@noop {} {\bibfield  {journal} {\bibinfo  {journal}
  {Annalen der Physik}\ }\textbf {\bibinfo {volume} {7}},\ \bibinfo {pages}
  {225} (\bibinfo {year} {1998})}\BibitemShut {NoStop}%
\bibitem [{\citenamefont {Furukawa}\ \emph {et~al.}(1998)\citenamefont
  {Furukawa}, \citenamefont {Rice},\ and\ \citenamefont
  {Salmhofer}}]{furukawa1998truncation}%
  \BibitemOpen
  \bibfield  {author} {\bibinfo {author} {\bibfnamefont {N.}~\bibnamefont
  {Furukawa}}, \bibinfo {author} {\bibfnamefont {T.}~\bibnamefont {Rice}}, \
  and\ \bibinfo {author} {\bibfnamefont {M.}~\bibnamefont {Salmhofer}},\
  }\href@noop {} {\bibfield  {journal} {\bibinfo  {journal} {Phys. Rev. Lett.}\
  }\textbf {\bibinfo {volume} {81}},\ \bibinfo {pages} {3195} (\bibinfo {year}
  {1998})}\BibitemShut {NoStop}%
\bibitem [{\citenamefont {Honerkamp}\ \emph {et~al.}(2001)\citenamefont
  {Honerkamp}, \citenamefont {Salmhofer}, \citenamefont {Furukawa},\ and\
  \citenamefont {Rice}}]{honerkamp2001breakdown}%
  \BibitemOpen
  \bibfield  {author} {\bibinfo {author} {\bibfnamefont {C.}~\bibnamefont
  {Honerkamp}}, \bibinfo {author} {\bibfnamefont {M.}~\bibnamefont
  {Salmhofer}}, \bibinfo {author} {\bibfnamefont {N.}~\bibnamefont {Furukawa}},
  \ and\ \bibinfo {author} {\bibfnamefont {T.~M.}\ \bibnamefont {Rice}},\
  }\href@noop {} {\bibfield  {journal} {\bibinfo  {journal} {Phys. Rev. B}\
  }\textbf {\bibinfo {volume} {63}},\ \bibinfo {pages} {035109} (\bibinfo
  {year} {2001})}\BibitemShut {NoStop}%
\bibitem [{\citenamefont {Nandkishore}\ \emph {et~al.}(2012)\citenamefont
  {Nandkishore}, \citenamefont {Levitov},\ and\ \citenamefont
  {Chubukov}}]{nandkishore2012chiral}%
  \BibitemOpen
  \bibfield  {author} {\bibinfo {author} {\bibfnamefont {R.}~\bibnamefont
  {Nandkishore}}, \bibinfo {author} {\bibfnamefont {L.}~\bibnamefont
  {Levitov}}, \ and\ \bibinfo {author} {\bibfnamefont {A.}~\bibnamefont
  {Chubukov}},\ }\href@noop {} {\bibfield  {journal} {\bibinfo  {journal} {Nat.
  Phys}\ }\textbf {\bibinfo {volume} {8}},\ \bibinfo {pages} {158} (\bibinfo
  {year} {2012})}\BibitemShut {NoStop}%
\bibitem [{Note1()}]{Note1}%
\bibitem [{\citenamefont {Li}\ \emph {et~al.}(2008)\citenamefont
  {Li}, \citenamefont {Checkelsky},\citenamefont {Hor},
  \citenamefont {Uher},\citenamefont {Hebard},\citenamefont {Cava},\ and\ \citenamefont
  {Ong}}]{Li2008}%
  \BibitemOpen
  \bibfield  {author} {\bibinfo {author} {\bibfnamefont {L.}~\bibnamefont
  {Li}}, \bibinfo {author} {\bibfnamefont {J.G.}~\bibnamefont
  {Checkelsky}}, \bibinfo {author} {\bibfnamefont {Y.S..}~\bibnamefont
  {Hor}}, \bibinfo {author} {\bibfnamefont {C.}~\bibnamefont
  {Uher}}, \bibinfo {author} {\bibfnamefont {A.F.}~\bibnamefont
  {Hebard}}, \bibinfo {author} {\bibfnamefont {R..J.}~\bibnamefont
  {Cava}}, \ and\ \bibinfo {author} {\bibfnamefont {N.P.}~\bibnamefont
  {Ong}},\ }\href@noop {} {\bibfield  {journal} {\bibinfo  {journal}
  {Science}\ }\textbf {\bibinfo {volume} {321}} (\bibinfo {year}
  {2008})}\BibitemShut {NoStop}%
  \bibitem [{\citenamefont {Mikitik}\ \emph {et~al.}(2015)\citenamefont
  {Mikitik}, \ and\ \citenamefont {Shirlai}}]{Mikitik2015}%
  \BibitemOpen
  \bibfield  {author} {\bibinfo {author} {\bibfnamefont {G.P.}~\bibnamefont
  {Mikitik}} \ and\ \bibinfo {author} {\bibfnamefont {Yu. V.}~\bibnamefont
  {Sharlai}},\ }\href@noop {} {\bibfield  {journal} {\bibinfo  {journal}
  {Phys. Rev. B}\ }\textbf {\bibinfo {volume} {91}} (\bibinfo {year}
  {2015})}\BibitemShut {NoStop}%
   \bibitem [{\citenamefont {Weller}\ \emph {et~al.}(2005)\citenamefont
  {Weller}, \citenamefont {Ellerby},
  \citenamefont {Saxena},\citenamefont {Smith},\ and \citenamefont {Skipper}}]{Weller2005}%
  \BibitemOpen
  \bibfield  {author} {\bibinfo {author} {\bibfnamefont {T.E.}~\bibnamefont
  {Weller}}, \bibinfo {author} {\bibfnamefont {M.}~\bibnamefont
  {Ellerby}}, \bibinfo {author} {\bibfnamefont {S.S.}~\bibnamefont
  {Saxena}}, \bibinfo {author} {\bibfnamefont {R.P.}~\bibnamefont
  {Smith}},\ and \bibinfo {author} {\bibfnamefont {N.T.}~\bibnamefont
  {Skipper}},\ }\href@noop {} {\bibfield  {journal} {\bibinfo  {journal}
  {Nat. Phys.}\ }\textbf {\bibinfo {volume} {1}} (\bibinfo {year}
  {2005})}\BibitemShut {NoStop}%
   \bibitem [{\citenamefont {Yoshioka}\ \emph {et~al.}(1981)\citenamefont
  {Yoshioka}, \ and\\citenamefont {Fukuyama}}]{Yoshioka1981}%
  \BibitemOpen
  \bibfield  {author} {\bibinfo {author} {\bibfnamefont {D.}~\bibnamefont
  {Yoshioka}}, \ and \bibinfo {author} {\bibfnamefont {H.}~\bibnamefont
  {Fukuyama}},\ }\href@noop {} {\bibfield  {journal} {\bibinfo  {journal}
  {J. Phys. Soc. Jpn}\ }\textbf {\bibinfo {volume} {50}} (\bibinfo {year}
  {1981})}\BibitemShut {NoStop}%
  \bibitem [{\citenamefont {Zhu}\ \emph {et~al.}(2018)\citenamefont
  {Zhu}, \citenamefont {Fauque},
  \citenamefont {Behnia},\ and\\citenamefont {Fuseya}}]{Zhu2018}%
  \BibitemOpen
  \bibfield  {author} {\bibinfo {author} {\bibfnamefont {Z.}~\bibnamefont
  {Zhu}}, \bibinfo {author} {\bibfnamefont {B.}~\bibnamefont
  {Fauque}}, \bibinfo {author} {\bibfnamefont {K.}~\bibnamefont
  {Behnia}}, \ and \bibinfo {author} {\bibfnamefont {Y.}~\bibnamefont
  {Fuseya}},\ }\href@noop {} {\bibfield  {journal} {\bibinfo  {journal}
  {J. Phys. Condens. Matter.}\ }\textbf {\bibinfo {volume} {30}} (\bibinfo {year}
  {2018})}\BibitemShut {NoStop}%
  \bibitem [{\citenamefont {Fujimori}\ \emph {et~al.}(2019)\citenamefont
  {Fujimori}, \citenamefont {Kawasaki},
  \citenamefont {Takeda},\citenamefont {Yamagami},\citenamefont {Nakamura},\citenamefont {Homma},\ and\\citenamefont {Aoki}}]{Fujimori2019}%
  \BibitemOpen
  \bibfield  {author} {\bibinfo {author} {\bibfnamefont {S.}~\bibnamefont
  {Fujimori}}, \bibinfo {author} {\bibfnamefont {I.}~\bibnamefont
  {Kawasaki}}, \bibinfo {author} {\bibfnamefont {Y.}~\bibnamefont
  {Takeda}}, \bibinfo {author} {\bibfnamefont {H.}~\bibnamefont
  {Yamagami}}, \bibinfo {author} {\bibfnamefont {A.}~\bibnamefont
  {Nakamura}}, \bibinfo {author} {\bibfnamefont {Y.}~\bibnamefont
  {Homma}}, \ and \bibinfo {author} {\bibfnamefont {D.}~\bibnamefont
  {Aoki}},\ }\href@noop {} {\bibfield  {journal} {\bibinfo  {journal}
  {J. Phys. Soc. Jpn}\ }\textbf {\bibinfo {volume} {88}} (\bibinfo {year}
  {2019})}\BibitemShut {NoStop}%
   \bibitem [{\citenamefont {Knebel}\ \emph {et~al.}(2019)\citenamefont
  {Knebel}, \citenamefont {Knafo},
  \citenamefont {Pourret},\citenamefont {Niu},\citenamefont {Valiska},\citenamefont {Braithwaite},\citenamefont {Lapertot},\citenamefont {Nardone},\citenamefont {Zitouni},\citenamefont {Mishra},\citenamefont {Sheikin},\citenamefont {Seyfarth},\citenamefont {Brison},\citenamefont {Aoki},\ and\\citenamefont {Flouquet}}]{Knebel2019}%
  \BibitemOpen
  \bibfield  {author} {\bibinfo {author} {\bibfnamefont {G.}~\bibnamefont
  {Knebel}}, \bibinfo {author} {\bibfnamefont {W.}~\bibnamefont
  {Knafo}}, \bibinfo {author} {\bibfnamefont {A.}~\bibnamefont
  {Pourret}}, \bibinfo {author} {\bibfnamefont {Q.}~\bibnamefont
  {Niu}}, \bibinfo {author} {\bibfnamefont {M.}~\bibnamefont
  {Valiska}}, \bibinfo {author} {\bibfnamefont {D.}~\bibnamefont
  {Braithwaite}}, \bibinfo {author} {\bibfnamefont {G.}~\bibnamefont
  {Lapertot}}, \bibinfo {author} {\bibfnamefont {M.}~\bibnamefont
  {Nardone}}, \bibinfo {author} {\bibfnamefont {A.}~\bibnamefont
  {Zitouni}}, \bibinfo {author} {\bibfnamefont {S.}~\bibnamefont
  {Mishra}}, \bibinfo {author} {\bibfnamefont {I.}~\bibnamefont
  {Sheikin}}, \bibinfo {author} {\bibfnamefont {G.}~\bibnamefont
  {Seyfarth}}, \bibinfo {author} {\bibfnamefont {J.P.}~\bibnamefont
  {Brison}}, \bibinfo {author} {\bibfnamefont {D.}~\bibnamefont
  {Aoki}}, \ and \bibinfo {author} {\bibfnamefont {J.}~\bibnamefont
  {Flouquet}},\ }\href@noop {} {\bibfield  {journal} {\bibinfo  {journal}
  {J. Phys. Soc. Jpn}\ }\textbf {\bibinfo {volume} {88}} (\bibinfo {year}
  {2019})}\BibitemShut {NoStop}%
  \bibitem [{\citenamefont {Ran}\ \emph {et~al.}(2019)\citenamefont
  {Ran}, \citenamefont {Liu},
  \citenamefont {Eo},\citenamefont {Campbell},\citenamefont {Neves},\citenamefont {Fuhrman},\citenamefont {Saha},\citenamefont {Eckberg},\citenamefont {Kim},\citenamefont {Graf},\citenamefont {Balakirev},\citenamefont {Singleton},\citenamefont {Paglione},\ and\\citenamefont {Butch}}]{Ran2019}%
  \BibitemOpen
  \bibfield  {author} {\bibinfo {author} {\bibfnamefont {S.}~\bibnamefont
  {Ran}}, \bibinfo {author} {\bibfnamefont {I-L.}~\bibnamefont
  {Liu}}, \bibinfo {author} {\bibfnamefont {Y.S.}~\bibnamefont
  {Eo}}, \bibinfo {author} {\bibfnamefont {D.J.}~\bibnamefont
  {Campbell}}, \bibinfo {author} {\bibfnamefont {P.M.}~\bibnamefont
  {Neves}}, \bibinfo {author} {\bibfnamefont {W.T.}~\bibnamefont
  {Fuhrman}}, \bibinfo {author} {\bibfnamefont {S.R.}~\bibnamefont
  {Saha}}, \bibinfo {author} {\bibfnamefont {C.}~\bibnamefont
  {Eckberg}}, \bibinfo {author} {\bibfnamefont {H.}~\bibnamefont
  {Kim}}, \bibinfo {author} {\bibfnamefont {D.}~\bibnamefont
  {Graf}}, \bibinfo {author} {\bibfnamefont {F.}~\bibnamefont
  {Balakirev}}, \bibinfo {author} {\bibfnamefont {J.}~\bibnamefont
  {Singleton}}, \bibinfo {author} {\bibfnamefont {J.}~\bibnamefont
  {Paglione}}, \ and \bibinfo {author} {\bibfnamefont {N.P.}~\bibnamefont
  {Butch}},\ }\href@noop {} {\bibfield  {journal} {\bibinfo  {journal}
  {Nat. Phys.}\ }\textbf {\bibinfo {volume} {15}} (\bibinfo {year}
  {2019})}\BibitemShut {NoStop}%
  \bibitem [{\citenamefont {Ran}\ \emph {et~al.}(2019)\citenamefont
  {Ran}, \citenamefont {Eckberg},
  \citenamefont {Ding},\citenamefont {Furukawa},\citenamefont {Metz},\citenamefont {Saha},\citenamefont {Liu},\citenamefont {Zic},\citenamefont {Kim},\citenamefont {Paglione},\ and\\citenamefont {Butch}}]{Ran684}%
  \BibitemOpen
  \bibfield  {author} {\bibinfo {author} {\bibfnamefont {S.}~\bibnamefont
  {Ran}}, \bibinfo {author} {\bibfnamefont {C.}~\bibnamefont
  {Eckberg}}, \bibinfo {author} {\bibfnamefont {Q-P.}~\bibnamefont
  {Ding}}, \bibinfo {author} {\bibfnamefont {Y.}~\bibnamefont
  {Furukawa}}, \bibinfo {author} {\bibfnamefont {T.}~\bibnamefont
  {Metz}},  \bibinfo {author} {\bibfnamefont {S.R.}~\bibnamefont
  {Saha}}, \bibinfo {author} {\bibfnamefont {I-L.}~\bibnamefont
  {Liu}}, \bibinfo {author} {\bibfnamefont {M.}~\bibnamefont
  {Zick}}, \bibinfo {author} {\bibfnamefont {H.}~\bibnamefont
  {Kim}}, \bibinfo {author} {\bibfnamefont {J.}~\bibnamefont
  {Paglione}},\ and \bibinfo {author} {\bibfnamefont {N.P.}~\bibnamefont
  {Butch}},\ }\href@noop {} {\bibfield  {journal} {\bibinfo  {journal}
  {Science}\ }\textbf {\bibinfo {volume} {365}} (\bibinfo {year}
  {2019})}\BibitemShut {NoStop}%
  \bibitem [{\citenamefont {Aoki}\ \emph {et~al.}(2019)\citenamefont
  {Aoki}, \citenamefont {Nakamura},
  \citenamefont {Honda},\citenamefont {Li},\citenamefont {Homma},\citenamefont {Shimizu},\citenamefont {Sato},\citenamefont {Knebel},\citenamefont {Brison},\citenamefont {Pourret},\citenamefont {Braithwaite},\citenamefont {Lapertot},\citenamefont {Niu},\citenamefont {Valiska},\citenamefont {Harima},\ and\\citenamefont {Flouquet}}]{Aoki2019}%
  \BibitemOpen
  \bibfield  {author} {\bibinfo {author} {\bibfnamefont {D.}~\bibnamefont
  {Aoki}}, \bibinfo {author} {\bibfnamefont {A.}~\bibnamefont
  {Nakamura}}, \bibinfo {author} {\bibfnamefont {F.}~\bibnamefont
  {Honda}}, \bibinfo {author} {\bibfnamefont {D.}~\bibnamefont
  {Li}}, \bibinfo {author} {\bibfnamefont {Y.}~\bibnamefont
  {Homma}}, \bibinfo {author} {\bibfnamefont {Y.}~\bibnamefont
  {Shimizu}}, \bibinfo {author} {\bibfnamefont {Y.J.}~\bibnamefont
  {Sato}}, \bibinfo {author} {\bibfnamefont {G.}~\bibnamefont
  {Knebel}}, \bibinfo {author} {\bibfnamefont {J-P.}~\bibnamefont
  {Brison}}, \bibinfo {author} {\bibfnamefont {A.}~\bibnamefont
  {Pourret}}, \bibinfo {author} {\bibfnamefont {D.}~\bibnamefont
  {Braithwaite}}, \bibinfo {author} {\bibfnamefont {G.}~\bibnamefont
  {Lapertot}}, \bibinfo {author} {\bibfnamefont {Q.}~\bibnamefont
  {Niu}}, bibinfo {author} {\bibfnamefont {M.}~\bibnamefont
  {Valiska}}, bibinfo {author} {\bibfnamefont {H.}~\bibnamefont
  {Harima}}, \ and \bibinfo {author} {\bibfnamefont {J.}~\bibnamefont
  {Flouquet}},\ }\href@noop {} {\bibfield  {journal} {\bibinfo  {journal}
  {J. Phys. Soc. Jpn}\ }\textbf {\bibinfo {volume} {88}} (\bibinfo {year}
  {2019})}\BibitemShut {NoStop}%
  \bibitem [{\citenamefont {Miyake}\ \emph {et~al.}(2019)\citenamefont
  {Miyake}, \citenamefont {Shimizu},
  \citenamefont {Sato},\citenamefont {Li},\citenamefont {Nakamura},\citenamefont {Homma},\citenamefont {Honda},\citenamefont {Flouquet},\citenamefont {Tokunaga},\ and\\citenamefont {Aoki}}]{Miyake2019}%
  \BibitemOpen
  \bibfield  {author} {\bibinfo {author} {\bibfnamefont {A.}~\bibnamefont
  {Miyake}}, \bibinfo {author} {\bibfnamefont {Y.}~\bibnamefont
  {Shimizu}}, \bibinfo {author} {\bibfnamefont {Y.J.}~\bibnamefont
  {Sato}}, \bibinfo {author} {\bibfnamefont {D.}~\bibnamefont
  {Li}}, \bibinfo {author} {\bibfnamefont {A.}~\bibnamefont
  {Nakamura}}, \bibinfo {author} {\bibfnamefont {Y.}~\bibnamefont
  {Homma}}, \bibinfo {author} {\bibfnamefont {F.}~\bibnamefont
  {Honda}}, \bibinfo {author} {\bibfnamefont {J.}~\bibnamefont
  {Flouquet}}, \bibinfo {author} {\bibfnamefont {M.}~\bibnamefont
  {Tokunaga}}, \ and \bibinfo {author} {\bibfnamefont {D.}~\bibnamefont
  {Aoki}},\ }\href@noop {} {\bibfield  {journal} {\bibinfo  {journal}
  {J. Phys. Soc. Jpn}\ }\textbf {\bibinfo {volume} {88}} (\bibinfo {year}
  {2019})}\BibitemShut {NoStop}%
    \bibitem [{\citenamefont {Tokunaga}\ \emph {et~al.}(2019)\citenamefont
  {Tokunaga}, \citenamefont {Sakai},
  \citenamefont {Kambe},\citenamefont {Hattori},\citenamefont {Higa},\citenamefont {Nakamine},\citenamefont {Kitagawa},\citenamefont {Ishida},\citenamefont {Nakamura},\citenamefont {Shimizu},\citenamefont {Homma},\citenamefont {Li},\citenamefont {Honda},\ and\\citenamefont {Aoki}}]{Tokunaga2019}%
  \BibitemOpen
  \bibfield  {author} {\bibinfo {author} {\bibfnamefont {Y.}~\bibnamefont
  {Tokunaga}}, \bibinfo {author} {\bibfnamefont {H.}~\bibnamefont
  {Sakai}}, \bibinfo {author} {\bibfnamefont {S.}~\bibnamefont
  {Kambe}}, \bibinfo {author} {\bibfnamefont {T.}~\bibnamefont
  {Hattori}}, \bibinfo {author} {\bibfnamefont {N.}~\bibnamefont
  {Higa}}, \bibinfo {author} {\bibfnamefont {G.}~\bibnamefont
  {Nakamine}}, \bibinfo {author} {\bibfnamefont {S.}~\bibnamefont
  {Kitagawa}}, \bibinfo {author} {\bibfnamefont {K.}~\bibnamefont
  {Ishida}}, \bibinfo {author} {\bibfnamefont {A.}~\bibnamefont
  {Nakamura}}, \bibinfo {author} {\bibfnamefont {Y.}~\bibnamefont
  {Shimizu}}, \bibinfo {author} {\bibfnamefont {Y.}~\bibnamefont
  {Homma}}, \bibinfo {author} {\bibfnamefont {D.}~\bibnamefont
  {Li}}, \bibinfo {author} {\bibfnamefont {F.}~\bibnamefont
  {Honda}}, \ and \bibinfo {author} {\bibfnamefont {D.}~\bibnamefont
  {Aoki}},\ }\href@noop {} {\bibfield  {journal} {\bibinfo  {journal}
  {J. Phys. Soc. Jpn}\ }\textbf {\bibinfo {volume} {88}} (\bibinfo {year}
  {2019})}\BibitemShut {NoStop}%
   \bibitem [{\citenamefont {Sundar}\ \emph {et~al.}(2019)\citenamefont
  {Sundar}, \citenamefont {Gheidi},
  \citenamefont {Akintola},\citenamefont {Kote},\citenamefont {Dunsiger},\citenamefont {Ran},\citenamefont {Butch},\citenamefont {Saha},\citenamefont {Paglione},\citenamefont {Shimizu},\ and\\citenamefont {Sonier}}]{Sundar2019}%
  \BibitemOpen
  \bibfield  {author} {\bibinfo {author} {\bibfnamefont {S.}~\bibnamefont
  {Sundar}}, \bibinfo {author} {\bibfnamefont {S.}~\bibnamefont
  {Gheidi}}, \bibinfo {author} {\bibfnamefont {K.}~\bibnamefont
  {Akintola}}, \bibinfo {author} {\bibfnamefont {A.M.}~\bibnamefont
  {Cote}}, \bibinfo {author} {\bibfnamefont {S.R.}~\bibnamefont
  {Dunsiger}}, \bibinfo {author} {\bibfnamefont {S.}~\bibnamefont
  {Ran}}, \bibinfo {author} {\bibfnamefont {N.P.}~\bibnamefont
  {Butch}}, \bibinfo {author} {\bibfnamefont {S.R.}~\bibnamefont
  {Saha}}, \bibinfo {author} {\bibfnamefont {J.}~\bibnamefont
  {Paglione}},\ and \bibinfo {author} {\bibfnamefont {J.E.}~\bibnamefont
  {Sonier}},\ }\href@noop {} {\bibfield  {journal} {\bibinfo  {journal}
  {Phys. Rev. B}\ }\textbf {\bibinfo {volume} {100}} (\bibinfo {year}
  {2019})}\BibitemShut {NoStop}%
   \bibitem [{\citenamefont {Metz}\ \emph {et~al.}(2019)\citenamefont
  {Metz}, \citenamefont {Bae},
  \citenamefont {Ran},\citenamefont {Liu},\citenamefont {Eo},\citenamefont {Fuhrman},\citenamefont {Agterberg},\citenamefont {Anlage},\citenamefont {Butch},\ and\\citenamefont {Paglione}}]{Metz2019}%
  \BibitemOpen
  \bibfield  {author} {\bibinfo {author} {\bibfnamefont {T.}~\bibnamefont
  {Metz}}, \bibinfo {author} {\bibfnamefont {S.}~\bibnamefont
  {Bae}}, \bibinfo {author} {\bibfnamefont {S.}~\bibnamefont
  {Ran}}, \bibinfo {author} {\bibfnamefont {I-L.}~\bibnamefont
  {Liu}}, \bibinfo {author} {\bibfnamefont {Y.S.}~\bibnamefont
  {Eo}}, \bibinfo {author} {\bibfnamefont {W.T.}~\bibnamefont
  {Fuhrman}}, \bibinfo {author} {\bibfnamefont {D.S.}~\bibnamefont
  {Agterberg}}, \bibinfo {author} {\bibfnamefont {S.M.}~\bibnamefont
  {Anlage}}, \bibinfo {author} {\bibfnamefont {N.P.}~\bibnamefont
  {Butch}}, \ and \bibinfo {author} {\bibfnamefont {J.}~\bibnamefont
  {Paglione}},\ }\href@noop {} {\bibfield  {journal} {\bibinfo  {journal}
  {Phys. Rev. B}\ }\textbf {\bibinfo {volume} {100}} (\bibinfo {year}
  {2019})}\BibitemShut {NoStop}%
  \bibitem [{\citenamefont {Ishizuka}\ \emph {et~al.}(2019)\citenamefont
  {Ishizuka}, \citenamefont {Sumita},
  \citenamefont {Daido},\ and\\citenamefont {Yanase}}]{Ishizuka2019}%
  \BibitemOpen
  \bibfield  {author} {\bibinfo {author} {\bibfnamefont {J.}~\bibnamefont
  {Ishizuka}}, \bibinfo {author} {\bibfnamefont {S.}~\bibnamefont
  {Sumita}}, \bibinfo {author} {\bibfnamefont {A.}~\bibnamefont
  {Daido}}, \ and \bibinfo {author} {\bibfnamefont {Y.}~\bibnamefont
  {Yanase}},\ }\href@noop {} {\bibfield  {journal} {\bibinfo  {journal}
  {Phys. Rev. Lett.}\ }\textbf {\bibinfo {volume} {123}} (\bibinfo {year}
  {2019})}\BibitemShut {NoStop}%
  \bibitem [{\citenamefont {Braithwaite}\ \emph {et~al.}(2019)\citenamefont
  {Braithwaite}, \citenamefont {Valiska},
  \citenamefont {Knebel},\citenamefont {Lapertot},\citenamefont {Brison},\citenamefont {Pourret},\citenamefont {Zhitomirsky},\citenamefont {Flouquet},\citenamefont {Honda},\ and\\citenamefont {Aoki}}]{Braithwaite2019}%
  \BibitemOpen
  \bibfield  {author} {\bibinfo {author} {\bibfnamefont {D.}~\bibnamefont
  {Braithwaite}}, \bibinfo {author} {\bibfnamefont {M.}~\bibnamefont
  {Valiska}}, \bibinfo {author} {\bibfnamefont {G.}~\bibnamefont
  {Knebel}}, \bibinfo {author} {\bibfnamefont {G.}~\bibnamefont
  {Lapertot}}, \bibinfo {author} {\bibfnamefont {J.-P.}~\bibnamefont
  {Brison}}, \bibinfo {author} {\bibfnamefont {A.}~\bibnamefont
  {Pourret}}, \bibinfo {author} {\bibfnamefont {M.E.}~\bibnamefont
  {Zhitomirsky}}, \bibinfo {author} {\bibfnamefont {J.}~\bibnamefont
  {Flouquet}}, \bibinfo {author} {\bibfnamefont {F.}~\bibnamefont
  {Honda}}, \ and \bibinfo {author} {\bibfnamefont {D.}~\bibnamefont
  {Aoki}},\ }\href@noop {} {\bibfield  {journal} {\bibinfo  {journal}
  {Communications Physics}\ }\textbf {\bibinfo {volume} {2}} (\bibinfo {year}
  {2019})}\BibitemShut {NoStop}%
  \bibitem [{\citenamefont {Miao}\ \emph {et~al.}(2020)\citenamefont
  {Miao}, \citenamefont {Liu},
  \citenamefont {Xu},\citenamefont {Kotta},\citenamefont {Kang},\citenamefont {Ran},\citenamefont {Paglione},\citenamefont {Kotliar},\citenamefont {Butch},\citenamefont {Denlinger},\ and\\citenamefont {Wray}}]{Miao2020}%
  \BibitemOpen
  \bibfield  {author} {\bibinfo {author} {\bibfnamefont {L.}~\bibnamefont
  {Miao}}, \bibinfo {author} {\bibfnamefont {S.}~\bibnamefont
  {Liu}}, \bibinfo {author} {\bibfnamefont {Y.}~\bibnamefont
  {Xu}}, \bibinfo {author} {\bibfnamefont {E.}~\bibnamefont
  {Kotta}}, \bibinfo {author} {\bibfnamefont {C-J.}~\bibnamefont
  {Kang}}, \bibinfo {author} {\bibfnamefont {S.}~\bibnamefont
  {Ran}}, \bibinfo {author} {\bibfnamefont {J.}~\bibnamefont
  {Paglione}}, \bibinfo {author} {\bibfnamefont {G.}~\bibnamefont
  {Kotliar}}, \bibinfo {author} {\bibfnamefont {N.P.}~\bibnamefont
  {Butch}}, \bibinfo {author} {\bibfnamefont {J.D.}~\bibnamefont
  {Denlinger}}, \ and \bibinfo {author} {\bibfnamefont {A.L.}~\bibnamefont
  {Wray}},\ }\href@noop {} {\bibfield  {journal} {\bibinfo  {journal}
  {Phys. Rev. Lett.}\ }\textbf {\bibinfo {volume} {124}} (\bibinfo {year}
  {2020})}\BibitemShut {NoStop}%
    \bibitem [{\citenamefont {Niu}\ \emph {et~al.}(2019)\citenamefont
  {Niu}, \citenamefont {Knebel},
  \citenamefont {Braithwaite},\citenamefont {Aoki},\citenamefont {Lapertot},\citenamefont {Seyfarth},\citenamefont {Brison},\citenamefont {Flouquet},\ and\\citenamefont {Pourret}}]{Niu2019}%
  \BibitemOpen
  \bibfield  {author} {\bibinfo {author} {\bibfnamefont {Q.}~\bibnamefont
  {Niu}}, \bibinfo {author} {\bibfnamefont {G.}~\bibnamefont
  {Knebel}}, \bibinfo {author} {\bibfnamefont {D.}~\bibnamefont
  {Braithwaite}}, \bibinfo {author} {\bibfnamefont {D.}~\bibnamefont
  {Aoki}}, \bibinfo {author} {\bibfnamefont {G.}~\bibnamefont
  {Lapertot}}, \bibinfo {author} {\bibfnamefont {G.}~\bibnamefont
  {Seyfarth}}, \bibinfo {author} {\bibfnamefont {J-P.}~\bibnamefont
  {Brison}}, \bibinfo {author} {\bibfnamefont {J.}~\bibnamefont
  {Flouquet}}, \ and \bibinfo {author} {\bibfnamefont {A.}~\bibnamefont
  {Pourret}},\ }\href@noop {} {\bibfield  {journal} {\bibinfo  {journal}
  {arXiv:1907.11118}\ }(\bibinfo {year}
  {2019})}\BibitemShut {NoStop}%
    \bibitem [{\citenamefont {Hutanu}\ \emph {et~al.}(2019)\citenamefont
  {Hutanu}, \citenamefont {Deng},
  \citenamefont {Ran},\citenamefont {Fuhrman},\citenamefont {Thoma},\citenamefont {Butch},\citenamefont {Paglione},\citenamefont {Kotliar},\ and\\citenamefont {Butch}}]{Hutanu2019}%
  \BibitemOpen
  \bibfield  {author} {\bibinfo {author} {\bibfnamefont {V.}~\bibnamefont
  {Hutanu}}, \bibinfo {author} {\bibfnamefont {H.}~\bibnamefont
  {Deng}}, \bibinfo {author} {\bibfnamefont {S.}~\bibnamefont
  {Ran}}, \bibinfo {author} {\bibfnamefont {W.T.}~\bibnamefont
  {Fuhrman}}, \bibinfo {author} {\bibfnamefont {H.}~\bibnamefont
  {Thoma}}, \ and \bibinfo {author} {\bibfnamefont {N.P.}~\bibnamefont
  {Butch}},\ }\href@noop {} {\bibfield  {journal} {\bibinfo  {journal}
  {arXiv:1905.04377}\ }(\bibinfo {year}
  {2019})}\BibitemShut {NoStop}%
    \bibitem [{\citenamefont {Jiao}\ \emph {et~al.}(2019)\citenamefont
  {Jiao}, \citenamefont {Howard},
  \citenamefont {Ran},\citenamefont {Wang},\citenamefont {Rodriguez},\citenamefont {Sigrist},\citenamefont {Wang},\citenamefont {Butch},\ and\\citenamefont {Madhavan}}]{Jiao2019}%
  \BibitemOpen
  \bibfield  {author} {\bibinfo {author} {\bibfnamefont {L.}~\bibnamefont
  {Jiao}}, \bibinfo {author} {\bibfnamefont {S.}~\bibnamefont
  {Howard}}, \bibinfo {author} {\bibfnamefont {S.}~\bibnamefont
  {Ran}}, \bibinfo {author} {\bibfnamefont {Z.}~\bibnamefont
  {Wang}}, \bibinfo {author} {\bibfnamefont {J.O.}~\bibnamefont
  {Rodriguez}}, \bibinfo {author} {\bibfnamefont {M.}~\bibnamefont
  {Sigrist}}, \bibinfo {author} {\bibfnamefont {Z.}~\bibnamefont
  {Wang}}, \bibinfo {author} {\bibfnamefont {N.}~\bibnamefont
  {Butch}}, \ and \bibinfo {author} {\bibfnamefont {V.}~\bibnamefont
  {Madhavan}},\ }\href@noop {} {\bibfield  {journal} {\bibinfo  {journal}
  {arXiv:1908.02846}\ }(\bibinfo {year}
  {2019})}\BibitemShut {NoStop}%
    \bibitem [{\citenamefont {Ran}\ \emph {et~al.}(2019)\citenamefont
  {Ran}, \citenamefont {Kim},
  \citenamefont {Liu},\citenamefont {Saha},\citenamefont {Hayes},\citenamefont {Metz},\citenamefont {Eo},\citenamefont {Paglione},\\ and\\citenamefont {Butch}}]{Rann2019}%
  \BibitemOpen
  \bibfield  {author} {\bibinfo {author} {\bibfnamefont {S.}~\bibnamefont
  {Ran}}, \bibinfo {author} {\bibfnamefont {H.}~\bibnamefont
  {Kim}}, \bibinfo {author} {\bibfnamefont {I-L.}~\bibnamefont
  {Liu}}, \bibinfo {author} {\bibfnamefont {S.}~\bibnamefont
  {Saha}}, \bibinfo {author} {\bibfnamefont {I.}~\bibnamefont
  {Hayes}}, \bibinfo {author} {\bibfnamefont {T.}~\bibnamefont
  {Metz}}, \bibinfo {author} {\bibfnamefont {Y.S.}~\bibnamefont
  {Eo}}, \bibinfo {author} {\bibfnamefont {J.}~\bibnamefont
  {Paglione}}, \ and \bibinfo {author} {\bibfnamefont {N.P.}~\bibnamefont
  {Butch}},\ }\href@noop {} {\bibfield  {journal} {\bibinfo  {journal}
  {arXiv:1909.06932}\ }(\bibinfo {year}
  {2019})}\BibitemShut {NoStop}%
    \bibitem [{\citenamefont {Bae}\ \emph {et~al.}(2019)\citenamefont
  {Bae}, \citenamefont {Kim},
  \citenamefont {Ran},\citenamefont {Eo},\citenamefont {Liu},\citenamefont {Fuhrman},\citenamefont {Paglione},\citenamefont {Butch},\ and\\citenamefont {Anlage}}]{Bae2019}%
  \BibitemOpen
  \bibfield  {author} {\bibinfo {author} {\bibfnamefont {S.}~\bibnamefont
  {Bae}}, \bibinfo {author} {\bibfnamefont {H.}~\bibnamefont
  {Kim}}, \bibinfo {author} {\bibfnamefont {S.}~\bibnamefont
  {Ran}}, \bibinfo {author} {\bibfnamefont {Y.S.}~\bibnamefont
  {Eo}}, \bibinfo {author} {\bibfnamefont {I-L.}~\bibnamefont
  {Liu}}, \ and \bibinfo {author} {\bibfnamefont {S.}~\bibnamefont
  {Anlage}},\ }\href@noop {} {\bibfield  {journal} {\bibinfo  {journal}
  {arXiv:1909.09032}\ }(\bibinfo {year}
  {2019})}\BibitemShut {NoStop}%
    \bibitem [{\citenamefont {Yarzhemsky}\ \emph {et~al.}(2020)\citenamefont
  {Yarzhemsky}, \ and\\citenamefont {Teplyakov}}]{Yarzhemsky2020}%
  \BibitemOpen
  \bibfield  {author} {\bibinfo {author} {\bibfnamefont {V.G.}~\bibnamefont
  {Yarzhemsky}}, \ and \bibinfo {author} {\bibfnamefont {E.A.}~\bibnamefont
  {Teplyaskov}},\ }\href@noop {} {\bibfield  {journal} {\bibinfo  {journal}
  {arXiv:2001.02963}\ }(\bibinfo {year}
  {2019})}\BibitemShut {NoStop}%
\end{thebibliography}

\appendix

\newpage

\begin{widetext}

\section{Expressions for bosonic couplings in terms of fermionic couplings}\label{app:A}

The expressions for the $g_{\alpha}$'s in Eq.5 of the main text in terms of the fermionic
interactions $g_{i}^{(j)}$ in Eq.1 are given by-
\[
g_{1},g_{2},g_{3}\rightarrow\frac{4}{(2\pi a)^2}(g_{2}^{(1)}-g_{1}^{(2)})
\]
\[
g_{4},g_{5},g_{6}\rightarrow\frac{4}{(2\pi a)^2}(g_{3}^{(2)}-g_{3}^{(1)})
\]
\[
g_{7},g_{8},g_{9}\rightarrow\frac{4}{(2\pi a)^2}(g_{1}^{(3)}-g_{2}^{(3)})
\]

In the RG equations in Eq. 7 the expressions for $\Lambda_{1}$ and $\Lambda_{-1}$
are given by
\[
\Lambda_{\pm1}=\frac{1}{32\pi}((a_{1}^{2}+a_{-1}^{2})^{2} g_{1}^{2}+(b_{1}^{2}+b_{-1}^{2})^{2} g_{2}^{2}+(c_{1}^{2}+c_{-1}^{2})^{2} g_{3}^{2}\pm(a_{1}^{8}g_{1}^{4}+4a_{1}^{6}a_{-1}^{2}g_{1}^{4}+4a_{1}^{2}a_{-1}^{6}g_{1}^{4}+a_{-1}^{8}g_{1}^{4}
\]
\[
+(b_{1}^{2}+b_{-1}^{2})^{4}g_{2}^{4}+2(b_{1}^{2}+b_{-1}^{2})(c_{1}^{2}+c_{-1}^{2})(b_{-1}(-c_{1}+c_{-1})+b_{1}(c_{1}+c_{-1}))(b_{1}(c_{1}-c_{-1})+
\]
\[
b_{-1}(c_{1}+c_{-1}))g_{2}^{2}g_{3}^{2}+(c_{1}^{2}+c_{-1}^{2})^{4}g_{3}^{4}+8a_{1}^{3}a_{-1}g_{1}^{2}(b_{1}b_{-1}(b_{1}^{2}+b_{-1}^{2})g_{2}^{2}+c_{1}c_{-1}(c_{1}^{2}+c_{-1}^{2})g_{3}^{2})+
\]
\[
8a_{1}a_{-1}^{3}g_{1}^{2}(b_{1}b_{-1}(b_{1}^{2}+b_{-1}^{2})g_{2}^{2}+c_{1}c_{-1}(c_{1}^{2}+c_{-1}^{2})g_{3}^{2})+2a_{1}^{4}g_{1}^{2}(3a_{-1}^{4}g_{1}^{2}+
\]
\[
(b_{1}^{4}-b_{-1}^{4})g_{2}^{2}+(c_{1}^{4}-c_{-1}^{4})g_{3}^{2})+2a_{-1}^{4}g_{1}^{2}((-b_{1}^{4}+b_{-1}^{4})g_{2}^{2}+(-c_{1}^{4}+c_{-1}^{4})g_{3}^{2}))^{1/2})
\]

\section{Order parameters, bosonic representation, scaling analysis}\label{app:B}

In our analysis, the fermionic bilinears for the order parameters
are expressed in terms of Gell-Mann matrices, which are a set of eight
linearly independent 3\texttimes 3 traceless Hermitian matrices, given
by-
\[
\lambda^{1}=\left(\begin{array}{ccc}
0 & 1 & 0\\
1 & 0 & 0\\
0 & 0 & 0
\end{array}\right),\lambda^{2}=\left(\begin{array}{ccc}
0 & -i & 0\\
i & 0 & 0\\
0 & 0 & 0
\end{array}\right)
\]

\[
\lambda^{3}=\left(\begin{array}{ccc}
1 & 0 & 0\\
0 & -1 & 0\\
0 & 0 & 0
\end{array}\right),\lambda^{4}=\left(\begin{array}{ccc}
0 & 0 & 1\\
0 & 0 & 0\\
1 & 0 & 0
\end{array}\right)
\]
\[
\lambda^{5}=\left(\begin{array}{ccc}
0 & 0 & -i\\
0 & 0 & 0\\
i & 0 & 0
\end{array}\right),\lambda^{6}=\left(\begin{array}{ccc}
0 & 0 & 0\\
0 & 0 & 1\\
0 & 1 & 0
\end{array}\right)
\]
\[
\lambda^{7}=\left(\begin{array}{ccc}
0 & 0 & 0\\
0 & 0 & -i\\
0 & i & 0
\end{array}\right),\lambda^{8}=\frac{1}{\sqrt{3}}\left(\begin{array}{ccc}
1 & 0 & 0\\
0 & 1 & 0\\
0 & 0 & -2
\end{array}\right)
\]

Below, we list the expressions for the eighteen order parameters in
terms of the bosonic fields:
\begin{align}
Re[O_{ph}^{00}] & \propto(2\cos[\sqrt{2}\sqrt{\pi}\widetilde{\phi}_{1}]\sin[\frac{2\sqrt{\pi}\widetilde{\phi}_{-1}}{\sqrt{6}}+\frac{2\sqrt{\pi}\widetilde{\phi}_{0}}{\sqrt{3}}-2k_{F}x]+\sin[-\frac{4}{\sqrt{6}}\sqrt{\pi}\widetilde{\phi}_{-1}+\frac{2\sqrt{\pi}\widetilde{\phi}_{0}}{\sqrt{3}}-2k_{F}x])\nonumber \\
Re[O_{ph}^{10}] & \propto\sin[\sqrt{2}\sqrt{\pi}\widetilde{\theta}_{1}]\cos[\frac{2\sqrt{\pi}\widetilde{\phi}_{-1}}{\sqrt{6}}+\frac{2\sqrt{\pi}\widetilde{\phi}_{0}}{\sqrt{3}}-2k_{F}x]\nonumber \\
Re[O_{ph}^{20}] & \propto\cos[\sqrt{2}\sqrt{\pi}\widetilde{\theta}_{1}]\cos[\frac{2\sqrt{\pi}\widetilde{\phi}_{-1}}{\sqrt{6}}+\frac{2\sqrt{\pi}\widetilde{\phi}_{0}}{\sqrt{3}}-2k_{F}x]\nonumber \\
Re[O_{ph}^{30}] & \propto\sin[\sqrt{2}\sqrt{\pi}\widetilde{\phi}_{1}]\cos[\frac{2\sqrt{\pi}\widetilde{\phi}_{-1}}{\sqrt{6}}+\frac{2\sqrt{\pi}\widetilde{\phi}_{0}}{\sqrt{3}}-2k_{F}x]\nonumber \\
Re[O_{ph}^{40}] & \propto\sin[\frac{\sqrt{\pi}\widetilde{\theta}_{1}}{\sqrt{2}}+\frac{3\sqrt{\pi}\widetilde{\theta}_{-1}}{\sqrt{6}}]\cos[\frac{\sqrt{\pi}\widetilde{\phi}_{1}}{\sqrt{2}}-\frac{\sqrt{\pi}\widetilde{\phi}_{-1}}{\sqrt{6}}+\frac{2\sqrt{\pi}\widetilde{\phi}_{0}}{\sqrt{3}}-2k_{F}x]\nonumber \\
Re[O_{ph}^{50}] & \propto\cos[\frac{\sqrt{\pi}\widetilde{\theta}_{1}}{\sqrt{2}}+\frac{3\sqrt{\pi}\widetilde{\theta}_{-1}}{\sqrt{6}}]\cos[\frac{\sqrt{\pi}\widetilde{\phi}_{1}}{\sqrt{2}}-\frac{\sqrt{\pi}\widetilde{\phi}_{-1}}{\sqrt{6}}+\frac{2\sqrt{\pi}\widetilde{\phi}_{0}}{\sqrt{3}}-2k_{F}x]\nonumber \\
Re[O_{ph}^{60}] & \propto\sin[\frac{\sqrt{\pi}\widetilde{\theta}_{1}}{\sqrt{2}}-\frac{3\sqrt{\pi}\widetilde{\theta}_{-1}}{\sqrt{6}}]\cos[\frac{\sqrt{\pi}\widetilde{\phi}_{1}}{\sqrt{2}}+\frac{\sqrt{\pi}\widetilde{\phi}_{-1}}{\sqrt{6}}-\frac{2\sqrt{\pi}\widetilde{\phi}_{0}}{\sqrt{3}}+2k_{F}x]\nonumber \\
Re[O_{ph}^{70}] & \propto\cos[\frac{\sqrt{\pi}\widetilde{\theta}_{1}}{\sqrt{2}}-\frac{3\sqrt{\pi}\widetilde{\theta}_{-1}}{\sqrt{6}}]\cos[\frac{\sqrt{\pi}\widetilde{\phi}_{1}}{\sqrt{2}}+\frac{\sqrt{\pi}\widetilde{\phi}_{-1}}{\sqrt{6}}-\frac{2\sqrt{\pi}\widetilde{\phi}_{0}}{\sqrt{3}}+2k_{F}x]\nonumber \\
Re[O_{ph}^{80}] & \propto(\cos[\sqrt{2}\sqrt{\pi}\widetilde{\phi}_{1}]\sin[\frac{2\sqrt{\pi}\widetilde{\phi}_{-1}}{\sqrt{6}}+\frac{2\sqrt{\pi}\widetilde{\phi}_{0}}{\sqrt{3}}-2k_{F}x]-\sin[-\frac{4}{\sqrt{6}}\sqrt{\pi}\widetilde{\phi}_{-1}+\frac{2\sqrt{\pi}\widetilde{\phi}_{0}}{\sqrt{3}}-2k_{F}x])\nonumber \\
Re[O_{pp}^{00}] & \propto(2\cos[\sqrt{2}\sqrt{\pi}\widetilde{\theta}_{1}]\sin[\frac{2\sqrt{\pi}\widetilde{\theta}_{-1}}{\sqrt{6}}+\frac{2\sqrt{\pi}\widetilde{\theta}_{0}}{\sqrt{3}}]-\sin[\frac{4\sqrt{\pi}\widetilde{\theta}_{-1}}{\sqrt{6}}-\frac{2\sqrt{\pi}\widetilde{\theta}_{0}}{\sqrt{3}}])\nonumber \\
Re[O_{pp}^{10}] & \propto\sin[\sqrt{2}\sqrt{\pi}\widetilde{\phi}_{1}]\cos[\frac{2\sqrt{\pi}\widetilde{\theta}_{-1}}{\sqrt{6}}+\frac{2\sqrt{\pi}\widetilde{\theta}_{0}}{\sqrt{3}}]\nonumber \\
Re[O_{pp}^{20}] & \propto\cos[\sqrt{2}\sqrt{\pi}\widetilde{\phi}_{1}]\cos[\frac{2\sqrt{\pi}\widetilde{\theta}_{-1}}{\sqrt{6}}+\frac{2\sqrt{\pi}\widetilde{\theta}_{0}}{\sqrt{3}}]\nonumber \\
Re[O_{pp}^{30}] & \propto\sin[\sqrt{2}\sqrt{\pi}\widetilde{\theta}_{1}]\cos[\frac{2\sqrt{\pi}\widetilde{\theta}_{-1}}{\sqrt{6}}+\frac{2\sqrt{\pi}\widetilde{\theta}_{0}}{\sqrt{3}}]\nonumber \\
Re[O_{pp}^{40}] & \propto\sin[\frac{\sqrt{\pi}\widetilde{\phi}_{1}}{\sqrt{2}}+\frac{3\sqrt{\pi}\widetilde{\phi}_{-1}}{\sqrt{6}}]\cos[\frac{\sqrt{\pi}\widetilde{\theta}_{1}}{\sqrt{2}}-\frac{\sqrt{\pi}\widetilde{\theta}_{-1}}{\sqrt{6}}+\frac{2\sqrt{\pi}\widetilde{\theta}_{0}}{\sqrt{3}}]\nonumber \\
Re[O_{pp}^{50}] & \propto\cos[\frac{\sqrt{\pi}\widetilde{\phi}_{1}}{\sqrt{2}}+\frac{3\sqrt{\pi}\widetilde{\phi}_{-1}}{\sqrt{6}}]\cos[\frac{\sqrt{\pi}\widetilde{\theta}_{1}}{\sqrt{2}}-\frac{\sqrt{\pi}\widetilde{\theta}_{-1}}{\sqrt{6}}+\frac{2\sqrt{\pi}\widetilde{\theta}_{0}}{\sqrt{3}}]\nonumber \\
Re[O_{pp}^{60}] & \propto\sin[\frac{\sqrt{\pi}\widetilde{\phi}_{1}}{\sqrt{2}}-\frac{3\sqrt{\pi}\widetilde{\phi}_{-1}}{\sqrt{6}}]\cos[\frac{\sqrt{\pi}\widetilde{\theta}_{1}}{\sqrt{2}}+\frac{\sqrt{\pi}\widetilde{\theta}_{-1}}{\sqrt{6}}-\frac{2\sqrt{\pi}\widetilde{\theta}_{0}}{\sqrt{3}}]\nonumber \\
Re[O_{pp}^{70}] & \propto\cos[\frac{\sqrt{\pi}\widetilde{\phi}_{1}}{\sqrt{2}}-\frac{3\sqrt{\pi}\widetilde{\phi}_{-1}}{\sqrt{6}}]\cos[\frac{\sqrt{\pi}\widetilde{\theta}_{1}}{\sqrt{2}}+\frac{\sqrt{\pi}\widetilde{\theta}_{-1}}{\sqrt{6}}-\frac{2\sqrt{\pi}\widetilde{\theta}_{0}}{\sqrt{3}}]\nonumber \\
Re[O_{pp}^{80}] & \propto(\cos[\sqrt{2}\sqrt{\pi}\widetilde{\theta}_{1}]\sin[\frac{2\sqrt{\pi}\widetilde{\theta}_{-1}}{\sqrt{6}}+\frac{2\sqrt{\pi}\widetilde{\theta}_{0}}{\sqrt{3}}]+\sin[\frac{4\sqrt{\pi}\widetilde{\theta}_{-1}}{\sqrt{6}}-\frac{2\sqrt{\pi}\widetilde{\theta}_{0}}{\sqrt{3}}]).\label{eq:10}
\end{align}

We further define the order parameters
\begin{align}
Re[\Delta_{ph}^{10}] & \propto\sin[\sqrt{2}\sqrt{\pi}\widetilde{\phi_{1}}+\frac{2\sqrt{\pi}\widetilde{\phi_{-1}}}{\sqrt{6}}+\frac{2\sqrt{\pi}\widetilde{\phi_{0}}}{\sqrt{3}}-2k_{F}x]\nonumber \\
Re[\Delta_{ph}^{20}] & \propto\sin[\sqrt{2}\sqrt{\pi}\widetilde{\phi_{1}}-\frac{2\sqrt{\pi}\widetilde{\phi_{-1}}}{\sqrt{6}}-\frac{2\sqrt{\pi}\widetilde{\phi_{0}}}{\sqrt{3}}+2k_{F}x]\nonumber \\
Re[\Delta_{ph}^{30}] & \propto\sin[\frac{4\sqrt{\pi}\widetilde{\phi_{-1}}}{\sqrt{6}}-\frac{2\sqrt{\pi}\widetilde{\phi_{0}}}{\sqrt{3}}+2k_{F}x]\nonumber \\
Re[\Delta_{pp}^{10}] & \propto\sin[\sqrt{2}\sqrt{\pi}\widetilde{\theta_{1}}+\frac{2\sqrt{\pi}\widetilde{\theta_{-1}}}{\sqrt{6}}+\frac{2\sqrt{\pi}\widetilde{\theta_{0}}}{\sqrt{3}}]\nonumber \\
Re[\Delta_{pp}^{20}] & \propto\sin[\sqrt{2}\sqrt{\pi}\widetilde{\theta_{1}}-\frac{2\sqrt{\pi}\widetilde{\theta_{-1}}}{\sqrt{6}}-\frac{2\sqrt{\pi}\widetilde{\theta_{0}}}{\sqrt{3}}]\nonumber \\
Re[\Delta_{pp}^{30}] & \propto\sin[\frac{4\sqrt{\pi}\widetilde{\theta_{-1}}}{\sqrt{6}}-\frac{2\sqrt{\pi}\widetilde{\theta_{0}}}{\sqrt{3}}].\label{eq:18}
\end{align}
corresponding to particle-particle and particle-hole ordering on each of the three individual Fermi pockets, which we track in our RG analysis. 
The order parameters, introduced as infinitesimal test vertices, scale upon renormalization in the following manner:
\begin{align}
\frac{dO_{ph}^{10}}{dy} & =(2-\frac{1}{16\pi}((A_{1}^{(4)})^{2}+(A_{-1}^{(4)})^{2}+(a_{1}^{(7)})^{2}+(a_{-1}^{(7)})^{2})
+\frac{1}{16\pi}g_{4}((A_{1}^{(4)})^{2}+(A_{-1}^{(4)})^{2})O_{ph}^{10},
\nonumber \\
\frac{dO_{ph}^{20}}{dy} & =(2-\frac{1}{16\pi}((A_{1}^{(4)})^{2}+(A_{-1}^{(4)})^{2}+(a_{1}^{(7)})^{2}+(a_{-1}^{(7)})^{2})
-\frac{1}{16\pi}g_{4}((A_{1}^{(4)})^{2}+(A_{-1}^{(4)})^{2}))O_{ph}^{20},
\nonumber \\
\frac{dO_{ph}^{40}}{dy} & =(2-\frac{1}{16\pi}((a_{1}^{(9)})^{2}+(a_{-1}^{(9)})^{2}+(A_{1}^{(5)})^{2}+(A_{-1}^{(5)})^{2}))
+\frac{1}{16\pi}g_{5}((A_{1}^{(5)})^{2}+(A_{-1}^{(5)})^{2}))O_{ph}^{40},
\nonumber \\
\frac{dO_{ph}^{50}}{dy} & =(2-\frac{1}{16\pi}((a_{1}^{(9)})^{2}+(a_{-1}^{(9)})^{2}+(A_{1}^{(5)})^{2}+(A_{-1}^{(5)})^{2})
-\frac{1}{16\pi}g_{5}((A_{1}^{(5)})^{2}+(A_{-1}^{(5)})^{2}))O_{ph}^{50},
\nonumber \\
\frac{dO_{ph}^{60}}{dy} & =(2-\frac{1}{16\pi}((a_{1}^{(8)})^{2}+(a_{-1}^{(8)})^{2}+(A_{1}^{(6)})^{2}+(A_{-1}^{(6)})^{2}))
+\frac{1}{16\pi}g_{6}((A_{1}^{(6)})^{2}+(A_{-1}^{(6)})^{2}))O_{ph}^{60},
\nonumber \\
\frac{dO_{ph}^{70}}{dy} & =(2-\frac{1}{16\pi}((a_{1}^{(8)})^{2}+(a_{-1}^{(8)})^{2}+(A_{1}^{(6)})^{2}+(A_{-1}^{(6)})^{2})
-\frac{1}{16\pi}g_{6}((A_{1}^{(6)})^{2}+(A_{-1}^{(6)})^{2}))O_{ph}^{70},
\nonumber \\
\frac{d\Delta_{ph}^{10}}{dy} & =(2-\frac{1}{4\pi}(a{}_{1}^{2}+a_{-1}^{2}))\Delta_{ph}^{10},\nonumber \\
\frac{d\Delta_{ph}^{20}}{dy} & =(2-\frac{1}{4\pi}(b{}_{1}^{2}+b_{-1}^{2}))\Delta_{ph}^{20},\nonumber \\
\frac{d\Delta_{ph}^{30}}{dy} & =(2-\frac{1}{4\pi}(c{}_{1}^{2}+c_{-1}^{2}))\Delta_{ph}^{30},\nonumber \\
\frac{dO_{pp}^{10}}{dy} & =(2-\frac{1}{16\pi}((a_{1}^{(1)})^{2}+(a_{-1}^{(1)})^{2}+(A_{1}^{(7)})^{2}+(A_{-1}^{(7)})^{2})
+\frac{1}{16\pi}g_{1}((a_{1}^{(1)})^{2}+(a_{-1}^{(1)})^{2}))O_{pp}^{10},
\nonumber \\
\frac{dO_{pp}^{20}}{dy} & =(2-\frac{1}{16\pi}((a_{1}^{(1)})^{2}+(a_{-1}^{(1)})^{2}+(A_{1}^{(7)})^{2}+(A_{-1}^{(7)})^{2})
-\frac{1}{16\pi}g_{1}((a_{1}^{(1)})^{2}+(a_{-1}^{(1)})^{2}))O_{pp}^{20},
\nonumber \\
\frac{dO_{pp}^{40}}{dy} & =(2-\frac{1}{16\pi}((a_{1}^{(2)})^{2}+(a_{-1}^{(2)})^{2}+(A_{1}^{(9)})^{2}+(A_{-1}^{(9)})^{2})
+\frac{1}{16\pi}g_{2}((a_{1}^{(2)})^{2}+(a_{-1}^{(2)})^{2}))O_{pp}^{40},
\nonumber \\
\frac{dO_{pp}^{50}}{dy} & =(2-\frac{1}{16\pi}((a_{1}^{(2)})^{2}+(a_{-1}^{(2)})^{2}+(A_{1}^{(9)})^{2}+(A_{-1}^{(9)})^{2})
-\frac{1}{16\pi}g_{2}((a_{1}^{(2)})^{2}+(a_{-1}^{(2)})^{2}))O_{pp}^{50},
\nonumber \\
\frac{dO_{pp}^{60}}{dy} & =(2-\frac{1}{16\pi}((a_{1}^{(3)})^{2}+(a_{-1}^{(3)})^{2}+(A_{1}^{(8)})^{2}+(A_{-1}^{(8)})^{2})
+\frac{1}{16\pi}g_{3}((a_{1}^{(3)})^{2}+(a_{-1}^{(3)})^{2}))O_{pp}^{60},
\nonumber \\
\frac{dO_{pp}^{70}}{dy} & =(2-\frac{1}{16\pi}((a_{1}^{(3)})^{2}+(a_{-1}^{(3)})^{2}+(A_{1}^{(8)})^{2}+(A_{-1}^{(8)})^{2})
-\frac{1}{16\pi}g_{3}((a_{1}^{(3)})^{2}+(a_{-1}^{(3)})^{2}))O_{pp}^{70},
\nonumber \\
\frac{d\Delta_{pp}^{10}}{dy} & =(2-\frac{1}{4\pi}(A{}_{1}^{2}+A_{-1}^{2}))\Delta_{pp}^{10},\nonumber \\
\frac{d\Delta_{pp}^{20}}{dy} & =(2-\frac{1}{4\pi}(B{}_{1}^{2}+B_{-1}^{2}))\Delta_{pp}^{20},\nonumber \\
\frac{d\Delta_{pp}^{30}}{dy} & =(2-\frac{1}{4\pi}(C_{1}^{2}+C_{-1}^{2}))\Delta_{pp}^{30},\label{eq:17}
\end{align} 

\end{widetext}
where the coefficients of the fields $\tilde{\phi_{i}}$ and $\tilde{\theta_{i}}$ in the sine-Gordon terms representing the different order parameters $O_{ph}^{i0}$,$O_{pp}^{i0}$ are expressed in terms of $a_{i}^{(\alpha)},\alpha=1-3,7-9$ and $A_{i}^{(\alpha)},\alpha=1-3,7-9$ respectively. Note that we work in the regimes $K_{0}^{\phi}\gg K_{\bot}^{\phi}$ or $K_{0}^{\phi}\ll K_{\bot}^{\phi}$, which enables us to drop terms involving $K_{0}^{\phi}$ compared to those involving $K_{\bot}^{\phi}$,  in different regimes. 

Here $a_{i}^{(\alpha)},\alpha=1-3,7-9$ and $A_{i}^{(\alpha)},\alpha=4-6$ are the usual coefficients of the fields for the interaction couplings $g_{\alpha}$ as defined in Eq. \ref{eq:5-1} of the main text. On the other hand, the coefficients $A_{i}^{(\alpha)},\alpha=7-9$ are defined in the same way for the fields $\tilde{\theta_{i}}$ as $a_{i}^{(\alpha)},\alpha=7-9$ are defined for the fields $\tilde{\phi_{i}}$ in Eq. \ref{eq:5-1}. While the latter coefficients for the fields $\tilde{\theta_{i}}$ do not actually appear in the sine-Gordon terms corresponding to any of the interaction couplings considered by us in Eq. \ref{eq:5-1}, we introduce them here for simplicity, since the coefficients of the fields in some of the order parameters can be expressed neatly  in terms of these quantities. 

In addition, the coefficients of the fields for the sine-Gordon terms corresponding to the order parameters $\Delta_{pp/ph}^{i0}(i=1-3)$ (see Eq.\ref{eq:18} above) are defined as $a_{i},b_{i},c_{i}$ and $A_{i},B_{i},C_{i}$ (where $i=1,-1$), as these cannot be expressed in terms of the coefficients already defined for any of the interaction couplings $g_{\alpha}$ in Eq. \ref{eq:5-1}. 
Note that the scaling dimensions for these order parameters do not have any $O(g)$ corrections from any of the nine couplings $g_i,\,i=1-9$ considered by us. However there are $O(g)$ corrections to their scaling dimensions from other interaction terms with higher scaling dimensions that have been neglected in this analysis. The order parameters $\Delta_{pp/ph}^{i0}(i=1-3)$ simultaneously diverge in certain parameter regimes where $K_{\bot}^{\phi}$ takes extremely large or small values (depending on the type of order being considered) \textendash{} a manifestation of restoration of the $C_{3}$ symmetry that had been explicitly broken through the initial conditions of the RG. The specific nature of order in these enlarged symmetry phases requires consideration of higher order processes that couple the degenerate order parameters, and may have $s$-wave, $d$-wave or chiral $d$-wave symmetry (see main text for discussion).

\end{document}